\documentclass[showpacs,preprintnumbers,amsmath,amssymb,prb]{revtex4}

\usepackage{graphicx}% Include Fig.files
\usepackage{dcolumn}% Align table columns on decimal point
\usepackage{bm}% bold math
\usepackage{amsfonts}
\usepackage{amssymb}

\newcommand\etal{\mbox{\textit{et al.}}}

\newcommand{\no}[1]{}

\def\drawline#1#2{\raise 2.5pt\vbox{\hrule width #1pt height #2pt}}

\def\trian{\raise 1.25pt\hbox{$\scriptscriptstyle\triangle$}\nobreak\ }

\def\square{${\vcenter{\hrule height .4pt
        \hbox{\vrule width .4pt height 3pt \kern 3pt
        \vrule width .4pt}
        \hrule height .4pt}}$\nobreak\ }

\def\plus{\raise 1.25pt \hbox{$\scriptscriptstyle +$}\nobreak\ }

\def\etal{{\it et al. }}
                                     
\begin{document}

\title{Low-dimensional model of turbulent Rayleigh-B\'enard convection in a Cartesian cell with square domain}
\author{Jorge Bailon-Cuba and J\"org Schumacher}
\affiliation{Institut f\"ur Thermo- und Fluiddynamik, \\
                 Technische Universit\"at Ilmenau, \\
                  Postfach 100565, D-98693 Ilmenau, Germany}
\date{\today}                  

\begin{abstract}
A low-dimensional model (LDM) for turbulent Rayleigh-B\'enard convection in a Cartesian cell with square domain, based on the Galerkin projection of the Boussinesq equations onto a finite set of 
empirical eigenfunctions, is presented. The empirical eigenfunctions are obtained from a 
joint Proper Orthogonal Decomposition (POD) of the velocity and temperature fields using 
the Snapshot Method on the basis of a direct numerical simulation (DNS). The resulting LDM is a 
quadratic inhomogeneous system of coupled ordinary differential equations which we use to
describe the long-time temporal evolution of the large-scale mode amplitudes for a Rayleigh 
number of $10^5$ and a Prandtl number of 0.7.  The truncation to a finite number of degrees
of freedom, that does not exceed a number of 310 for the present,  requires the additional
implementation of an eddy viscosity-diffusivity to capture the missing dissipation of the
small-scale modes. The magnitude of this additional dissipation mechanism is determined
by requiring statistical stationarity and a total dissipation that corresponds with the original DNS 
data. We compare the performance of two models, a constant so-called Heisenberg viscosity--diffusivity and a mode-dependent or modal one. The latter viscosity--diffusivity model turns  out to reproduce 
the large-scale properties of the turbulent convection qualitatively well, even for a model with 
only a few hundred POD modes.
\end{abstract}

\maketitle

\section{Introduction}
For most turbulent flows in nature and technology, it is impossible to resolve all relevant 
degrees of freedom.  Systematic methods to derive models with a reduced number of degrees of
freedom from the full set of nonlinear fluid equations are thus necessary. Low-dimensional 
modeling of transient and turbulent flows using Galerkin projection onto the empirical basis 
functions which are obtained from a 
proper orthogonal decomposition (POD) is one such well established method. 
\cite{Lumley1971,berkooz,moehlis_2} POD and the development of Galerkin models based on 
POD modes has been applied to a number  of fundamental hydrodynamic flow problems, including simple wall-bounded shear flows, \cite{aubry,berkooz_2,moehlis_3,moehlis_1} flows over cavities 
\cite{cazemier,kalb} or in the wake of a cylinder.\cite{deane_1,ma_1,Noack2003,Rowley2006} 
The development of low-dimensional models (LDM) based on POD modes has also been 
extended  in several directions such as to the balanced POD method \cite{Rowley2005,Rowley2008} or to unsteady flow problems \cite{Noack2010} for which fast and slow flow modes are separated. 
Most of these cases have been studied for laminar or transitional flows at lower or moderate 
Reynolds numbers. 

With increasing Reynolds number the flows become turbulent, the number of degrees
of freedom grows rapidly and their nonlinear couplings are increasingly relevant. The 
truncation of the set of nonlinear ordinary differential equations (ODE) which follows from 
Galerkin projection introduces always a cut-off of these mode interactions and removes 
couplings between the degrees of freedom which are necessary for the transfer of
kinetic energy from large to small scales. An additional dissipation mechanism has to be 
implemented in the low-dimensional model to account for
the dominant dissipation by the truncated degrees of freedom. The particular way of truncation 
can then alter the dynamics in the LDM. 

Several approaches to this problem have been suggested in the past. 
Aubry \etal \cite{aubry} used directly the energy transfer between resolved modes and 
unresolved modes at smaller scales  to formulate a spectral closure in the truncated  system 
of ODEs. Moehlis \etal \cite{moehlis_3} presented streamwise-invariant truncations for a
plane Couette flow, and showed that very low-dimensional models with up to ten degrees 
of freedom can reproduce transient flow phenomena  in low-Reynolds-number shear flows. 
They also found that the detailed behavior of their LDM depends in a subtle manner on the 
modes included and that a proper account of the symmetries of the system is crucial. Later 
Smith \etal \cite{moehlis_1,moehlis_2} included streamwise variations 
in their model. They also introduced a linear damping term, but only 
when the particular POD mode expansion coefficient $a^{(n)}$ is significantly 
anti-correlated with its time derivative, $d a^{(n)}/dt$. Cazemier \etal \cite{cazemier} 
constructed a LDM for driven cavity flows, consisting of the 80 most energetic POD modes 
computed from 700 snapshots of a direct numerical simulation (DNS).
To study  the time evolution of the truncated ODE system, a slightly different linear damping term is introduced in their model. This term is calculated from the requirement that the energy of the 
ODE system is conserved in a statistically stationary sense.

A few attempts to derive LDMs are reported for Rayleigh-B\'{e}nard (RB) convection, 
despite being one of  the most comprehensively studied flows.\cite{Ahlers2009,Lohse2010} 
Studies of RB convection in a finite box, based on the POD procedure, have been mostly done 
by Sirovich and co-workers. \cite{sirovich_4,sirovich_5,sirovich_6,sirovich_2,sirovich_3} 
Sirovich and Park \cite{sirovich_5,sirovich_6} discussed the importance of the discrete symmetries describing the velocity--temperature fluctuations field. Deane and Sirovich \cite{sirovich_2} made a parametric study of the POD mode spectra for small Rayleigh numbers $Ra\le 46000$. 

Only recently, a snapshot method has been applied to turbulent RB convection in a closed
cylindrical cell for Rayleigh numbers up to $Ra=10^8$ and cell aspect ratios between one half and 
three.\cite{Bailon2010} In this work, emphasis was given to relating the first POD modes to the 
large-scale flow circulation which is always present in a closed turbulent convection cell. \cite{Ahlers2009} The disentanglement of the temperature and velocity fields into POD modes 
allowed the authors to quantify the amount of heat which is transported by the particular 
POD modes through the convection cell.  A change of the large-scale flow from a one-roll to 
a two-roll pattern, which is observed when the aspect ratio is increased beyond one at a 
fixed Rayleigh number, was in line with a decrease of transported heat by the primary mode 
compared to the secondary POD mode. 

As a correspondence of the few POD studies of RB convection, only a few works exist with an 
emphasis on developing a LDM by a Galerkin projection of the Boussinesq equations
onto the most energetic POD modes. Tarman \cite{tarman_1} derived a model from POD modes 
which have been however separately extracted from the velocity and temperature fields. In a second work he proposed an algorithm which incorporates the lost dissipation due to truncation.\cite{tarman_2} Besides the cutoff index based on the energy (mode index $k < k_e$), a second index based on the dissipation ($k_d > k_e$) was considered. The time dependence of the modes with indexes $k_e < k \le k_d$ was expressed as the quotient of the corresponding nonlinear and dissipation coefficient. No closed forms for the constant coefficients in the ODE system were however obtained in any of these works. 

In the present work, we want to extend these studies of RB convection in several directions.
First, we construct a LDM for the evolution of the POD mode coefficients $a^{(n)}$ in the 
case of turbulent Rayleigh-B\'{e}nard convection in a Cartesian cell with periodic side walls
and isothermal free-slip top and bottom square planes. It is essential to use POD modes of the combined 
four-vector velocity-temperature field.\cite{Bailon2010} In this derivation, it turns out that  a 
cubic term due to the interaction of the velocity with the mean temperature field 
(denoted as $\Im$) becomes linear as a consequence of the orthogonality of the POD modes. 
The other terms which arise in the Galerkin projection are a linear production term $\wp$,  a linear dissipation term $\epsilon$, a quadratic nonlinear term $N$, and a constant term $\epsilon_{\langle T\rangle}$ corresponding to the dissipation due to the mean temperature field. Second, our 
studies will extend previous works  \cite{tarman_2,sirovich_2,sirovich_6} in terms of the magnitude of 
the Rayleigh number of convection. A case with $Ra\sim 10^5$ is considered for which 
RB convection is turbulent and a DNS data record  exists. 
Third, we are interested in the long-time behavior of the dynamics in the LDM.  With a view to
more complex convection flows in the future, we are seeking for the least set of POD modes 
that can reproduce characteristic dynamics of turbulent convection.

A solution which includes the additional dissipation due to the neglected less energetic POD 
modes has to be considered by an additional eddy viscosity--diffusivity, $\eta \ge 0$. First, we 
present the so--called Heisenberg model with a constant $\eta$ which exerts the same fraction of dissipation on all POD modes. As will be shown, this closure requires at least a minimum number of degrees of freedom, in particular with respect to the vertical direction, for a qualitatively correct description of the flow. As a consequence, two LDMs with 210 and  310 degrees of freedom, respectively, are chosen. They are taken from a set of 15708 modes (see Sec. III B). As will be seen, this model fails to reproduce the large-scale evolution of convection.
For the larger of the two sets of modes, the model relaxes to a statistically stationary state which 
contains too much energy.  Second, we refine this model and include a mode--dependent (or modal) eddy viscosity--diffusivity. The magnitude of both eddy viscosity--diffusivity contributions has to be estimated. In order to do so, we will follow a procedure that has been suggested by Cazemier \etal \cite{cazemier}. The second model yields much more realistic large-scale variations of the most energetic modes, also reproducing with reasonable accuracy the energy spectrum and the turbulence statistics.
Therefore, a significant part of the present work discusses the impact of both types of eddy viscosity-diffusivity $\eta$ on the dynamics of the LDM with different number of degrees of freedom and how it compares to DNS. 

The outline of the paper is as follows. The equations of motion, the basic idea of POD -- in 
particular for the method of snapshots -- is discussed in the next section. The construction of the 
LDM by Galerkin projection onto POD modes of RB convection follows in Sec. III.
In this section, the results of the time integration of the LDM with both eddy viscosity--diffusivity schemes, and the agreement with the DNS are also discussed. We conclude with a summary and give an outlook.
 
\section{Methods}
\subsection{Equations of motion and numerical scheme}
\label{num_mod}
\begin{figure}[t]
        \centering
         \includegraphics[width=8.0cm]{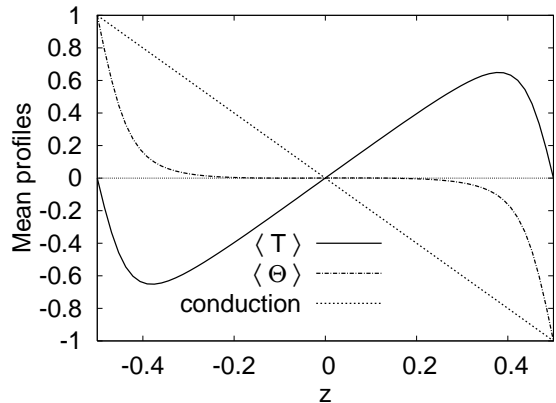}
\caption{Mean vertical profiles $\langle T(z)\rangle$, $\langle\Theta(z)\rangle$ and the 
linear conduction profile (see Eq. (\ref{decomposition})).  Data are obtained from the DNS
data record.}
\label{fig:temp_all}
\end{figure}

Turbulent Rayleigh-B\'{e}nard convection is governed by the Boussinesq 
equations. They are brought into a dimensionless form by rescaling with the domain height 
$l_z$, the diffusive time scale $t_{diff}=l_z^2/\kappa$  with $\kappa$ being the thermal diffusivity, the temperature difference $\Delta T=\Theta_{bottom}-\Theta_{top}>0$ and follow to
\begin{equation}
{\nabla}\cdot \mathbf{u}=0
\label{eq:continuity}
\end{equation}
\begin{equation} 
\frac{\partial{\mathbf{u}}}{\partial{t}}+(\mathbf{u}\cdot{\nabla})\mathbf{u}=-{\nabla}p+Pr{\nabla}^2\mathbf{u} + Ra Pr T \mathbf{e}_z \label{eq:NS}
\end{equation}
\begin{equation} 
\frac{\partial{T}}{\partial{t}}+(\mathbf{u}\cdot{\nabla})T = w + {\nabla}^2 T
\label{eq:energy}
\end{equation}
where $\mathbf{u}(\mathbf{x},t)$ is the velocity field, $T(\mathbf{x},t)$ the departure from the linear conduction temperature profile, and $p(\mathbf{x},t)$ is the kinematic pressure. Dimensionless 
parameters are the Rayleigh number $Ra=g \alpha \Delta T {l_z}^3/(\nu \kappa)$ and the Prandtl
number  $Pr= \nu/\kappa$. Besides diffusivity $\kappa$, they contain 
the kinematic viscosity $\nu$, the gravitational acceleration $g$, and the thermal expansion coefficient 
$\alpha$. Note that the total 
temperature field is given in our notation by (see also Ref. \cite{sirovich_5})
\begin{equation}
\Theta(\mathbf{x},t)=\Theta_{bottom}-\frac{\Delta T}{l_z} z +T(\mathbf{x},t)\,.
\label{decomposition}
\end{equation}
The vector $\mathbf{e}_z$ 
is the direction in which buoyancy and gravity work and in which the mean temperature gradient is established. The dimensions of the cell $\Omega$ are $l_x=l_y=4\pi, l_z=\pi$, 
where from now on $x, y$ are the horizontal and $z$ the vertical dimensionless 
coordinates. The aspect ratio is fixed to $l_x/l_z=l_y/l_z=4$. For convenience, the origin of the coordinate system is in the center of the cell. Therefore, $x\in [-L_x/2,L_x/2]$, 
$y\in [-L_y/2,L_y/2]$, and $z\in [-1/2,1/2]$ with $L_x=l_x/l_z$ and $L_y=l_y/l_z$. The $x$-, 
$y$- and $z$-components of the velocity field will be denoted by $u$, $v$ and $w$, respectively. 
The boundary conditions are periodic in $x$ and $y$, and free-slip in $z$. This means that at the 
hot bottom plane at  $z/l_z=-1/2$ and the cold top plane at $z/l_z=1/2$ the following conditions hold:
\begin{equation}
w= T = \frac{\partial u}{\partial z}= \frac{\partial v}{\partial z}= 0\,. 
\end{equation}
For the present boundary conditions, the flow can be decomposed in
\begin{subequations} 
\begin{align} 
        u(\mathbf{x},t)  =\langle u(z) \rangle + u'(\mathbf{x},t) \\
        v(\mathbf{x},t)  =\langle v(z) \rangle + v'(\mathbf{x},t)  \\   
        w(\mathbf{x},t)  =\langle w(z) \rangle + w'(\mathbf{x},t) \\
        T(\mathbf{x},t)  =\langle T(z) \rangle + \theta(\mathbf{x},t) 
\end{align}
\end{subequations}
where the mean components, e.g. $\langle u(z)\rangle$ are ensemble averages obtained by 
averaging over the horizontal $x$-$y$ plane and time, i.e., a sequence of $N_T'$ statistically 
independent snapshots. The ensemble average for the velocity component $u$ is thus given by
\begin{equation}
\langle u(z)\rangle =\frac{1}{N_T'} \sum_{n=1}^{N_T'}\Biggl( \frac{1}{L_x L_y} \int_{-L_x/2}^{L_x/2} \int_{-L_y/2}^{L_y/2} u(x,y,z,t_n)\,dy\,dx\Biggr)\,.
\label{eq:mean_avg}\end{equation}
Figure \ref{fig:temp_all} shows the mean vertical profiles of $\Theta$ and $T$ together with the linear thermal conduction profile.  
%-----------------------------------------------------------------
%-----------------------------------------------------------------
Considering the fact that for our problem, due to symmetry considerations, $\langle u \rangle =\langle v \rangle =\langle w \rangle = 0$, we can take the four-vector field
\begin{equation}
\mathbf{v}(\mathbf{x},t)= (u,v,w,\theta)
\end{equation}
for the POD analysis and LDM  derivation. Our analysis is in the statistically stationary regime
of convective turbulence. The ensemble average of~(\ref{eq:energy}) yields then an expression linking the mean and the fluctuating components of the flow in the form
\begin{equation}
\frac {d^2 \langle T \rangle}{dz^2}=\frac {d\langle w \theta \rangle}{dz}
\label{eq:eavg_energy}
\end{equation}
Following Sirovich \etal\cite{sirovich_4} temperatures and velocities are rescaled with  
$\hat{\theta}_c = \sqrt{2 Nu^3/Ra \ Pr}$ and $\hat{u}_c = \sqrt{Ra \ Pr/2Nu}$, respectively. 
Here $Nu$ is the Nusselt number. The quantity $\hat{\theta}_c$ is obtained by demanding that 
the turbulent heat flux in the center of the cell must be equal to that due to diffusion at 
the boundary.

The highly-resolved  data record is obtained by a pseudo-spectral DNS which uses fast Fourier 
transformations.\cite{Schumacher2008,Schumacher2009} Time stepping is done by a second-order
Runge Kutta scheme. For most of the work, we consider 
a data set with $N_T'=320$ full three-dimensional turbulence snapshots which are separated by 
four convective time units $t_{conv}=\sqrt{l_z/g\alpha\Delta T}$ from each other. The 
computational grid consists of  $N_x\times N_y\times N_z=256 \times 256 \times 65$ points 
in $x$-, $y$-, and $z$-directions, respectively. The spectral resolution is given by $k_{max}
\eta_K=5.6$. Here, $k_{max}=\sqrt{2}N_z/3$
and $\eta_K$ the Kolmogorov dissipation length. The Rayleigh number is  $Ra=1.03 \times 10^5$ and the Prandtl number $Pr=0.7$.

\subsection{Proper Orthogonal Decomposition (POD)}
The POD is a model reduction technique that extracts the most energetic modes from a set of realizations or snapshots of the flow. These POD modes are used as a basis for Galerkin projections 
of the full set of nonlinear equations thus reducing the infinite-dimensional space of solutions 
to a finite-dimensional system.\cite{berkooz, sirovich} The two-point correlation tensor or 
covariance matrix of the four-vector field is defined by
\begin{equation}
K_{mn}(\mathbf{x},\mathbf{x'})= \langle v_m(\mathbf{x},t)v^*_n(\mathbf{x'},t)\rangle_t
\label{eq:two-point-corr}
\end{equation}
where the asterisk denotes the complex conjugate, $\langle\cdot\rangle_t$ the time average and $m,n=1,2,3,4$. For Rayleigh-B\'enard convection in Cartesian domains with two homogeneous 
(invariant with respect to translations) directions Eq.~(\ref{eq:two-point-corr}) takes the form \cite{Holmes_1}
\begin{equation}
K_{mn}(\mathbf{x},\mathbf{x'})= K_{mn}(x-x',y-y',z)
\label{eq:homog_xz}
\end{equation}
For a kernel~(\ref{eq:homog_xz}), the eigenfunctions have the form
\begin{equation}
\Phi^{(p)}_{m;n_x,n_y}(x,y,z)= \frac {\phi^{(p)}_{m;n_x,n_y}(z)}{\sqrt{L_x L_y}} \exp \biggr( \frac{2\pi i n_x x}{L_x}+\frac{2\pi i n_y y}{L_y} \biggr)
\label{eq:eigenfunc}
\end{equation}
where $n_x,n_y$ are integers for the $x-$and $y-$directions, respectively. 
The superscript $(p)$ denotes a particular POD mode.
The determination of $\phi$ follows then from
\begin{equation}
\sum_{n=1}^4 \int_{-1/2}^{1/2}\kappa_{mn}(n_x,n_y;z,z') \phi^{(p)}_{n;n_x,n_y}(z')\,dz' =\lambda^{(p)}_{n_x,n_y} \phi^{(p)}_{m;n_x,n_y}(z)
\label{eq:phi_eq}
\end{equation}
where $\kappa_{mn}$ is the Fourier transform of $K_{mn}$ with respect to the homogeneous directions
$x$ and $y$.  The kernel $\kappa_{mn}$ is calculated from the numerical data set by first taking the discrete Fourier transform of each realization in the horizontal plane,
\begin{equation}
F_m(n_x,n_y;z,t) = \frac {1}{L_x L_y} \int_{-L_x/2}^{L_x/2} \int_{-L_y/2}^{L_y/2} v_m(x,y,z,t)\exp \biggr[-\biggr(\frac{2\pi i n_x x}{L_x}+\frac{2\pi i n_y y}{L_y} \biggr)\biggr]\,dx \,dy
\label{eq:fourier_transf}
\end{equation}
and then averaging the correlation over the entire ensemble of data,
\begin{equation}
\kappa_{mn}(n_x,n_y;z,z')= \langle F_m(n_x,n_y;z,t)F^{*}_n(n_x,n_y;z',t)\rangle_t\,.
\label{eq:two-pt-fourier}
\end{equation}
Thus the kernel $\kappa_{mn}$ is Hermitian, non-negative and on physical grounds square integrable, such that the existence of a complete set of vector eigenfunctions $\{\phi^{(p)}_{m;n_x,n_y}(z)\}_{p=1\dots}$ given by~(\ref{eq:phi_eq}) is assured. Complex conjugation of Eq.~(\ref{eq:phi_eq}) and use of 
Eq.~(\ref{eq:eigenfunc}) implies that
\begin{equation}
\phi^{(p)}_{m;n_x,n_y}(z)= \phi^{(p)*}_{m;-n_x,-n_y}(z)\,.
\label{eq:eigenfunc-2}
\end{equation}
Due to the reality of the physical space fields $v_m(\mathbf{x},t)$ Eq. (\ref{eq:fourier_transf}) implies that
\begin{equation}
F^{*}_m(n_x,n_y;z,t)= F_m(-n_x,-n_y;z,t)\,.
\label{eq:fourier-coeff-2}
\end{equation}
The associated expansion of the velocity field $v_m(\mathbf{x},t)$ in terms of the modes is given as
\begin{equation}
v_m(\mathbf{x},t)= \sum_{p}\sum_{n_x}\sum_{n_y} \frac {a^{(p)}_{n_x,n_y}(t)} {\sqrt{L_x L_y}} \exp \biggr( \frac{2\pi i n_x x}{L_x}+\frac{2\pi i n_y y}{L_y} \biggr) \phi^{(p)}_{m;n_x,n_y}(z)\,,
\label{eq:vel-field-exp}
\end{equation}
and again reality of the four-vector field implies that $a^{(p)}_{n_x,n_y}(t)= a^{(p)*}_{-n_x,-n_y}(t)$. The index
$(p)$ runs over the POD modes. The coefficients are calculated by the scalar product in L$_2(\Omega)$, 
\begin{eqnarray}
a^{(p)}_{n_x,n_y}(t)&=& (\Phi^{(p)}_{m;n_x,n_y}(x,y,z),v_m(x,y,z,t))\\ \nonumber
                                   &=&
\sum_{m=1}^4 \int_{-L_x/2}^{L_x/2}\int_{-L_y/2}^{L_y/2}\int_{-1/2}^{1/2} \Phi^{(p)*}_{m;n_x,n_y}(x,y,z) v_m(x,y,z,t)\,dx\,dy\,dz\,.
\label{eq:dot-prod-int}
\end{eqnarray} 
Next, the discrete Fourier transform of $v_m(\mathbf{x},t)$ is introduced together with the ansatz
of $\Phi$ in (\ref{eq:eigenfunc}) and the orthogonality of the complex exponentials. This gives
\begin{equation}
a_{n_x,n_y}^{(p)}(t)= \sqrt{L_x L_y}\,\sum_{m=1}^4 \int_{-1/2}^{1/2} \phi^{(p)*}_{m;n_x,n_y}(z)F_m(n_x,n_y;z,t)\,dz\,.
\label{eq:dotprod-int_final}
\end{equation} 

\subsection{The method of snapshots}
The snapshot method is one way to obtain the POD modes, particularly when the computational 
grid size becomes large. The time coordinate $t$ in the equations above has to be substituted 
now by an index that runs over the sequence of snapshots. It is based on the fact that 
(\ref{eq:two-pt-fourier}) is a degenerate kernel.\cite{sirovich_4} Consequently, an eigenfunction of the kernel $\kappa_{mn}$ can be represented as
\begin{eqnarray}
\phi^{(p)}_{m;n_x,n_y}(z)& = &\sum_{l=1}^{N_T'} \sum_{\gamma\in{\cal G}} \gamma \cdot \alpha^{(p)}(n_x,n_y,l) \gamma \cdot F_m(n_x,n_y;z,l) \,,\nonumber\\
&=& \sum_{k=1}^{N_T}\alpha^{(p)}(n_x,n_y,k) F_m(n_x,n_y;z,k)\,,
\label{eq:snapshot}
\end{eqnarray}
where $\gamma$ as an element of the symmetry group ${\cal G}=\{\mathbf{G}, \mathbf{Z} \mathbf{G}\}$ as discussed in Appendix~\ref{app_1}. 
The explicit use of symmetries in the problem at hand enlarges the data record with originally
$N_T'$ snapshots to a total number of  $N_T=16 N_T'$ snapshots thus improving the 
convergence. For $N_T'=320$ DNS snapshots we thus end up with $N_T=5120$ samples that 
can be used to evaluate the POD modes. 
Replacing the kernel in~(\ref{eq:phi_eq}) and using Eq. (\ref{eq:snapshot}) results to
\begin{eqnarray}
& &\frac{L_x L_y}{N_T}\sum_{m=1}^{N_T}\biggr(\int_{-1/2}^{1/2} \sum_{j=1}^4 F^*_j(n_x,n_y;z',k)F_j(n_x,n_y;z',m)\,dz' \biggl)\alpha^{(p)}(n_x,n_y,m)\nonumber\\ 
&=&\lambda^{(p)}_{n_x,n_y}\,\alpha^{(p)}(n_x,n_y,k)\,,
\label{eq:snapshot_eq}
\end{eqnarray}
where $k,m=1,2,...,N_T$ represent any two snapshots (including all possible symmetries). 
Then (\ref{eq:snapshot_eq}) is the matrix problem which yields the 
eigenvalues $\lambda$ and eigenfunctions $\phi$. It is clear that it determines just $N_T$ of the empirical eigenfunctions for  a fixed tupel $(n_x, n_y)$. The eigenvalue $\lambda^{(p)}_{n_x,n_y}$ of 
the $N_T\times N_T$ matrix is the total energy (kinetic energy plus temperature variance) of the 
$p$th POD mode for $(n_x, n_y)$. Recall that in the present  case the four-velocity field is expanded into 
Fourier modes with respect to $x$ and $y$ which is characterized by wavenumbers $n_x$ and 
$n_y$, respectively. This results theoretically in an infinite set of POD modes. 

\section{Results}
\subsection{Galerkin Projection of the Boussinesq equations onto the POD modes}
\label{galerkin}
Given the full set of nonlinear Boussinesq equations and the POD modes extracted by a snapshot
method from the DNS data, we can proceed to derive the LDM. This requires first a
Galerkin projection step.  Using the dimensionless units introduced in 
section~\ref{num_mod}, Eqns.~(\ref{eq:NS}) and~(\ref{eq:energy}) can be rewritten together in a
four-vector notation with respect to $\mathbf{v}$,
\begin{align} 
\frac{\partial{v_i}}{\partial{t}}= -\hat{u}_c \sum_{j=1}^3 \ v_j \frac{\partial{v_i}}{\partial{x_j}}+ \frac {Pr}{1+\delta_{i4}(Pr-1)} \sum_{j=1}^3\frac{{\partial}^2{v_i}}{\partial{x^2_j}}+ 2 \delta_{i3}\,Nu^2 \, v_4 + \nonumber \\
\delta_{i4} \ \frac{\hat{u}_c}{\hat{\theta}_c} \ v_3 + \delta_{i4} \biggl( \sum_{j=1}^3\frac {{\partial}^2{\langle T \rangle}}{\partial{x^2_j}} - \hat{u}_c \sum_{j=1}^3 v_j \frac{\partial{\langle T \rangle}}{\partial{x_j}} -s_i \biggr) +s_i
 \label{eq:Tensor_NS_Bousinesq}
\end{align}
where $i=1,2,3,4$. Here, $j=1,2,3$ correspond to the three spatial coordinates and the term
\begin{equation}
s_i= 2 \delta_{i3} \, Nu^2 \ \langle T \rangle - \frac{1}{\hat{u}_c} \frac {\partial{p}}{\partial{x_i}}\,,
\end{equation}
is a source which drops out in the Galerkin projection procedure. This is due to the divergence-free 
nature of the POD basis functions and the fact that modes $\Phi^{(p)}_{3;0,0}(x,y,z)=0$, $\forall p$, 
respectively. Now one takes the inner product of (\ref{eq:Tensor_NS_Bousinesq}) with modes $\Phi^{(p)}_{m;p_x,p_y}(x,y,z)$ and inserts the expansion
$v_m (\mathbf{x},t)= \sum_{n}\sum_{n_x}\sum_{n_y} a^{(n)}_{n_x,n_y}(t) \Phi^{(n)}_{m;n_x,n_y}(x,y,z)$ from equation~(\ref{eq:vel-field-exp}). Due to the orthogonality of the POD modes, the following 
infinite-dimensional ODE system follows
\begin{align}
\dot{a}^{(p)}_{p_x,p_y} = \sum_{n=1}^{\infty} {\wp}(p_x,p_y,n,p) a^{(n)}_{p_x,p_y} + \sum_{n=1}^{\infty} {\epsilon}(p_x,p_y,n,p) a^{(n)}_{p_x,p_y} + \nonumber \\
{N}(p_x,p_y,p) + \sum_{n=1}^{\infty} \Im (p_x,p_y,n,p) a^{(n)}_{p_x,p_y} + {\epsilon_{\langle T \rangle}}(p_x,p_y,p)\,.
\label{eq:galerkin-proj}
\end{align}
The terms on the right hand side of (\ref{eq:galerkin-proj}) correspond to the production ($\wp$), dissipation ($\epsilon$), nonlinear transfer ($N$), interaction with the mean flow ($\Im$), and the dissipation due to the mean temperature field ($\epsilon_{\langle T \rangle}$), respectively.
Closed forms for $\epsilon$ and $N$ have been obtained in Refs.~\cite{sirovich_4} and~\cite{moehlis_2}, but only for the velocity field. The general equations for the terms on the right hand 
side of (\ref{eq:galerkin-proj}) are as follows. The production term is given by
\begin{equation}
{\wp}(p_x,p_y,n,p)= \int_{-1/2}^{1/2} \biggl(2 Nu^2\, \phi^{(n)}_{4;p_x,p_y} \phi^{(p)*}_{3;p_x,p_y} + \frac{\hat{u}_c}{\hat{\theta}_c}\ \phi^{(n)}_{3;p_x,p_y} \phi^{(p)*}_{4;p_x,p_y} \biggr) dz
\label{eq:production}
\end{equation} 
and the dissipation term  by
\begin{align}
{\epsilon}(p_x,p_y,n,p)= -Pr\sum_{j=1}^3 \biggl\{\biggl[\biggl(\frac{2\pi p_x}{L_x}\biggr)^2 + \biggl(\frac{2\pi p_y}{L_y}\biggr)^2 \biggr]\int_{-1/2}^{1/2} \phi^{(n)}_{j;p_x,p_y} \phi^{(p)*}_{j;p_x,p_y} dz + \nonumber \\  
\int_{-1/2}^{1/2} \phi^{(n)'}_{j;p_x,p_y} \phi^{(p)*'}_{j;p_x,p_y} dz \biggr\} - \nonumber \\
\biggl\{\biggl[\biggl(\frac{2\pi p_x}{L_x}\biggr)^2 + \biggl(\frac{2\pi p_y}{L_y}\biggr)^2 \biggr]\int_{-1/2}^{1/2} \phi^{(n)}_{4;p_x,p_y} \phi^{(p)*}_{4;p_x,p_y} dz + \nonumber \\  
\int_{-1/2}^{1/2} \phi^{(n)'}_{4;p_x,p_y} \phi^{(p)*'}_{4;p_x,p_y} dz \biggr\}\,.
\label{eq:dissipation}
\end{align} 
The nonlinear mode coupling term  $N$ is given by
\begin{align}
{N}(p_x,p_y,p)= \sum_{n=1}^{\infty}\sum_{m=1}^{\infty}\sum_{m_x=-\infty}^{\infty}\sum_{m_y=-\infty}^{\infty} {\hat{B}}^{(m,n,p)}(m_x,m_y,p_x,p_y) a^{(m)}_{m_x,m_y} a^{(n)}_{p_x-m_x, p_y-m_y}\,,
\label{eq:nonlinear}
\end{align} 
where the coefficients are given by
\begin{align}
{\hat{B}}^{(m,n,p)}(m_x,m_y,p_x,p_y)= - \frac {\hat{u}_c}{\sqrt{L_x L_y}} \sum_{j=1}^4\int_{-1/2}^{1/2} \biggl[\frac{2\pi i m_x}{L_x} \phi^{(n)}_{1;p_x-m_x,p_y-m_y} \phi^{(m)}_{j;m_x,m_y}+ \nonumber \\
\frac{2\pi i m_y}{L_y} \phi^{(n)}_{2;p_x-m_x,p_y-m_y} \phi^{(m)}_{j;m_x,m_y}+ \phi^{(n)}_{3;p_x-m_x,p_y-m_y} \phi^{(m)'}_{j;m_x,m_y} \biggr] \phi^{(p)*}_{j;p_x,p_y} dz\,.
\label{eq:nonlinear_2}
\end{align} 
The last two terms in Eq. (\ref{eq:galerkin-proj}), $\Im$ and ${\epsilon_{\langle T \rangle}}$, are calculated for a quasi-steady flow \cite{tarman_1}. 
The term $\Im$  is related to the ensemble average of temperature, $\mathbf{u}\cdot{\nabla} \langle T \rangle = w \frac{d \langle T \rangle}{dz}$. However, from the ensemble average of~(\ref{eq:energy}) we obtain
\begin{equation}
\frac{d \langle T \rangle}{dz}(z)=\hat{u}_c \langle w \theta (z)\rangle + \frac {d \langle T \rangle }{dz}(z=-1/2)
\label{eq:integral_T}
\end{equation}
where
$\frac {d \langle T \rangle }{dz}(z=-1/2)= - \hat{u}_c \int_{-1/2}^{1/2} \langle w \theta \rangle \ dz $.
In terms of the eigenvectors, 
\begin{equation}
\langle w \theta \rangle = \frac {1}{L_x L_y}\sum_{k=1}^{\infty}\sum_{l=1}^{\infty}\sum_{k_x=-\infty}^{\infty}\sum_{k_y=-\infty}^{\infty} \langle a^{(k)}_{k_x,k_y}(t) a^{(l)}_{-k_x,-k_y}(t) \rangle \,\phi^{(k)}_{3;k_x,k_y} \phi^{(l)}_{4;-k_x,-k_y}
\label{eq:shear_stress}
\end{equation}
since $\langle w \theta \rangle$ is only a function of $z$. However $\langle a^{(k)}_{k_x,k_y}(t) a^{(l)}_{-k_x,-k_y}(t) \rangle = \langle a^{(k)}_{k_x,k_y}(t) a^{(l)*}_{k_x,k_y}(t) \rangle = \delta_{kl} \lambda^{(k)}_{k_x,k_y}$ and therefore
\begin{eqnarray}
\langle w \theta \rangle &=& \frac {1}{L_x L_y} \sum_{k=1}^{\infty}\sum_{l=1}^{\infty}\sum_{k_x=-\infty}^{\infty}\sum_{k_y=-\infty}^{\infty} \delta_{kl} \lambda^{(k)}_{k_x,k_y} \phi^{(k)}_{3;k_x,k_y} \phi^{(l)}_{4;-k_x,-k_y} \\
&=& \frac {1}{L_x L_y} \sum_{k=1}^{\infty}\sum_{k_x=-\infty}^{\infty}\sum_{k_y=-\infty}^{\infty} \lambda^{(k)}_{k_x,k_y} \phi^{(k)}_{3;k_x,k_y} \phi^{(k)*}_{4;k_x,k_y}\,.  
\label{eq:shear_II}
\end{eqnarray}
Consequently, the equation for the term $\Im$ is
\begin{equation}
\Im(p_x,p_y,n,p)= - {\hat{u}_c}^2 \int_{-1/2}^{1/2} \biggl( \langle w \theta \rangle - \int_{-1/2}^{1/2} \langle w \theta \rangle \ dz \biggr)\ \phi^{(n)}_{3;p_x,p_y} \phi^{(p)*}_{4;p_x,p_y}\ dz 
\label{eq:interaction_II}\end{equation} 
where $\langle w \theta \rangle$ is given by (\ref{eq:shear_II}).  
For the low-dimensional description of free-shear-flow, Rajaee \etal \cite{mojtaba} keeps the original time dependence in (\ref{eq:shear_II}), just replacing $\lambda^{(k)}_{k_x,k_y} = \langle | a^{(k)}_{k_x,k_y} |^2 \rangle$ and considering this as a running-time-average factor, so the resulting $\sum_{n=1}^{\infty} \Im (k_x,k_y,n,k) a^{(n)}_{k_x,k_y}$ term becomes cubic in the ODE system (\ref{eq:galerkin-proj}).

Finally, we have a dissipation term which is related to the mean temperature profile, 
${\nabla}^2 \langle T \rangle$. The term is nonzero only for the purely 
thermal and real modes $\phi^{(n)}_{4;0,0}$. In the present case this corresponds to a maximum energy of 
$\lambda^{(3)}_{0,0}=0.1989\%$. The term is given by
\begin{equation}
\epsilon_{\langle T \rangle}(p_x,p_y,p) = \frac {\hat{u}_c} {\sqrt{L_x L_y}} \sum_{k=1}^{\infty}\sum_{k_x=-\infty}^{\infty}\sum_{k_y=-\infty}^{\infty} \lambda^{(k)}_{k_x,k_y} \int_{-1/2}^{1/2} ( \phi^{(k)}_{3;k_x,k_y} \phi^{(k)*}_{4;k_x,k_y} )^{'} \ \phi^{(p)}_{4;0,0} \ dz
\label{eq:mean_diss_II}
\end{equation}
only if $p_{x}=p_{y}=0$, otherwise $\epsilon_{\langle T \rangle}(p_x,p_y,p)=0$.
In Eq.~(\ref{eq:mean_diss_II}), we used again $(\cdot)^{'}=d (\cdot )/dz$. 
This completes the discussion of the different terms 
that arise due to the Galerkin projection of the Boussinesq equations onto the POD modes. 

%-----------------------------------------------------------------
\begin{table}
\begin{center}
\begin{tabular}{lcccc}
\hline\hline
Model  & $M$ &$n_x$ and $n_y$  & $(n)$  &  \% Energy \\
\hline
M1  & 210  & $\;\;\;|n_x|+|n_y|\le 4$ &   $\;\;\;1\le (n)\le 10$  &  76.894  \\
M2  & 310  & $\;\;\;|n_x|+|n_y|\le 5$ &   $\;\;\;1\le (n)\le 10$  &  82.202  \\
Total & 15708 & $\;\;\;\;0\le n_x\le 16\,,$ &  $\;\;\;1\le (n)\le 28$ &  100   \\
 &  & $-16\le n_y\le 16$ &    &      \\
\hline\hline
\end{tabular}  
\end{center}
\caption{Parameters of the two LDMs denoted as M1 and M2. We list the range 
of the horizontal wavenumbers $n_x$ and $n_y$ and the so-called quantum number 
$(n)$. In the last two lines of the table, the total number of POD modes that 
has been calculated is given. It follows from the maximum range of horizontal wave and 
vertical quantum numbers. The horizontal wavenumber in $x$-direction starts from 
zero due to symmetry of Eq. (\ref{eq:eigenfunc-2}). The total number of POD modes from
the snapshot analysis is thus  $M=17\times 33\times 28=15708$.}
\label{POD_wavenumbers}
\end{table}

%---------------------------------------------------------------------------
\begin{figure}[htpb!]
\centering
\includegraphics[width=8.0cm]{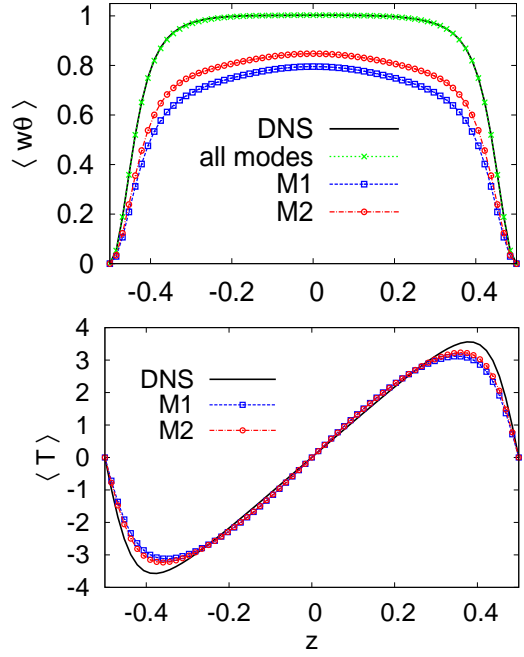}
\caption{(Color online) Reconstruction of the vertical profiles of mean convective heat 
flux (top) and temperature deviation from the linear profile (bottom). Data from DNS are 
compared with the two LDMs, M1 \& M2, as well as with the complete set of POD modes 
$(M=15708)$ obtained from the snapshot analysis.}
\label{fig:temperature}
\end{figure}
%------------------------------------------------------------------
\begin{figure}[htpb!]
\centering
\includegraphics[width=8.0cm]{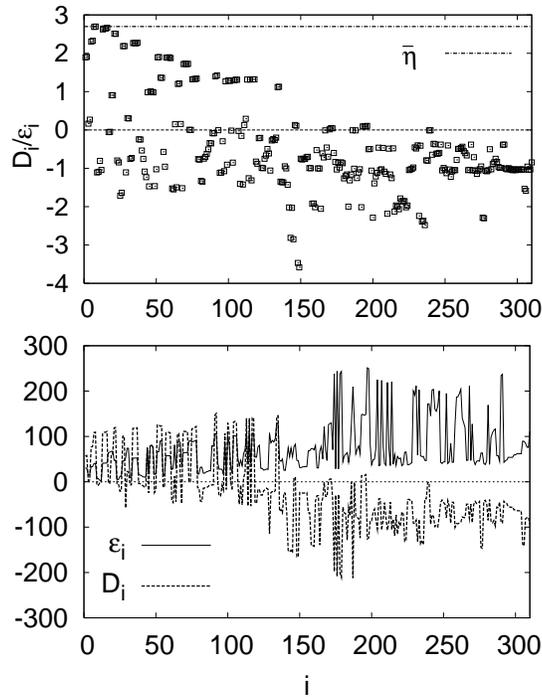}
\caption{Estimation of the maximum eddy viscosity and diffusivity following the procedure of 
Ref. 8. (Top) Ratio  $D_i/\epsilon_i$ for the LDM model M2 with 310 POD modes. 
The dashed line marks the maximum of the ratio. (Bottom) $D_i$ and $\epsilon_i$ are separately 
shown for the same data.}
\label{fig:cazemier_aubry}
\end{figure}
%-----------------------------------------------------------------

\subsection{Truncation to a low-dimensional model}
\label{truncation_LDM}
The integrals along the $z$--axis contained in the coefficients of Eq. (\ref{eq:galerkin-proj}) are evaluated on the computational grid of the DNS. Since the integrands are discrete functions of 
the wave and quantum numbers, we include all degrees of freedom with $0 \le n_x \le 16$, $-16 
\le n_y \le 16$, and $1 \le (n) \le 28$. This results in a maximum number of POD modes of 
$M=15708$. Out of this set of POD modes, we select small subsets of the most energetic POD 
modes which corresponds to the large-scale structures of the convection dynamics. 

The choice of $M=210$ for our first LDM denoted as M1 follows from the restriction to modes with 
$|n_x|+|n_y|\le 4$ and $1\le (n)\le 10$ as indicated in Tab.\ref{POD_wavenumbers}. This level 
of truncation builds on experiences from similar studies by Aubry \etal \cite{aubry}, 
Moehlis \etal \cite{moehlis_3}, Holmes \etal \cite{Holmes_1, Holmes_2} and Podvin \cite{podvin_2} 
in wall-bounded shear flows. They reproduced successfully the dynamics close to the walls 
for a range of values of the so-called Heisenberg parameter $\eta$ (which will be discussed below). Modes with $n_x=0$ in streamwise  and $-5 \le n_y \le 5$ in spanwise direction were taken. As 
explained by Holmes \etal \cite{Holmes_2}, this choice is due to the fact that for the given spanwise 
domain length, the cross--stream interactions that contribute to the observed bursts of the velocity 
are well reproduced with at least five nonzero modes.
Moreover, when we choose small quantum numbers, $(n) \le 5 $, the present LDM for convection yields 
solutions which decay monotonically with time. Our model and the resulting degeneracy restrictions for the average field--modes $\lambda^{(n)}_{0,0}$ did not allow us to take $(n) < 8$. A significant improvement of the LDM dynamics is obtained for $(n) \le 10$. This is due to the fact that the 
two additional average field--modes with $\lambda^{(9)}_{0,0}=\lambda^{(10)}_{0,0}$ have a degeneracy of 2 and will be incorporated. A second LDM called M2 was introduced with range of horizontal wave 
numbers $|n_x|+|n_y|\le 5$ (see Tab.~\ref{POD_wavenumbers}). The latter LDM will be used 
for most of the following studies.

Figure~\ref{fig:temperature} represents the Reynolds shear stress $\langle w \theta(z) \rangle$ 
and the average temperature profile $\langle T (z)\rangle$ as reconstructed from the two LDMs. 
The calculation of both profiles is done by integration of Eqns. (\ref{eq:shear_II}) and 
(\ref{eq:integral_T}), respectively and explained in section~\ref{galerkin}. As can be seen in 
the figure, the convergence to the vertical DNS profile for $\langle w \theta(z) \rangle$ is very slow. 
Only the significant enhancement of the degrees of freedom up to $M=15708$ results in an excellent agreement with the DNS profiles, for the present Rayleigh and Prandtl numbers.

The modes of the LDMs are however not the modes which contribute dominantly to thermal 
and kinetic energy dissipation. The missing couplings to the small-scale dissipating modes 
causes numerical stability problems of the LDM. Various methods have been proposed therefore  
to stabilize the truncated low-dimensional dynamical system as we have discussed in the 
introduction. Our studies showed that the model introduced by Cazemier \etal \cite{cazemier} worked best.
This model introduces a closure based on the mean energy balance as derived from Eq. 
(\ref{eq:galerkin-proj}) which can be rewritten as 
\begin{align}
\dot{a}^{(p)}_{p_x,p_y} = \sum_{n=1}^{M_Q} {A} (p_x,p_y,n,p) a^{(n)}_{p_x,p_y} + {N}(p_x,p_y,p) + 
\epsilon_{\langle T\rangle}(p_x,p_y,p)\,.
\label{eq:galerkin-cazemier}
\end{align}
where $M_Q$ is the highest quantum number in the LDM ($M_Q \le 28$).
The three linear terms in Eq. (\ref{eq:galerkin-proj}) are summarized to
\begin{align}
 {A} (p_x,p_y,n,p)= {\wp}(p_x,p_y,n,p)+ {\epsilon}(p_x,p_y,n,p)+ \Im (p_x,p_y,n,p)\,.
\label{eq:galerkin-linear}
\end{align}
From Eq. (\ref{eq:galerkin-cazemier}), one can derive an equation for the total energy by 
multiplication with $a^{(n)}_{p_x,p_y}$ and summation over all quantum and wavenumbers.
An additional linear damping term, ${D} (p_x,p_y,p)$, is then quantitatively determined 
from the requirement that the mean total energy of the extended dynamical system is in a 
statistically stationary state, i.e., 
\begin{eqnarray}             
0&=& \sum_{n,m,m_x,m_y} {\hat{B}}^{(m,n,p)}(m_x,m_y,p_x,p_y) \langle a^{(m)}_{m_x,m_y} a^{(n)}_{p_x-m_x, p_y-m_y} a^{(p)*}_{p_x,p_y} \rangle\nonumber\\ 
&+& 
\biggl( {A} (p_x,p_y,p,p) + {D} (p_x,p_y,p) \biggr) \langle a^{p}_{p_x,p_y} a^{p\,\ast}_{p_x,p_y}\rangle
+ 
\epsilon_{\langle T\rangle}(p_x,p_y,p) \langle a^{(p)*}_{p_x,p_y} \rangle \,.
\label{eq:cazemier_2}
\end{eqnarray}
Note, that in this real equation, the last term on the right hand side is nonzero only for the purely 
thermal modes with $p_x=p_y=0$. Also, due to orthogonality, the last two indices of $A$ have to be equal.

In the so-called Heisenberg dissipation model by Aubry \etal \cite{aubry} the action of the 
neglected modes on the ones contained in the LDM is represented in an average sense, 
namely  as a function of the dynamics of these coherent structures. An even simpler approach 
was chosen by Omurtag and Sirovich. \cite{omurtag} They simply introduced a constant empirical viscosity coefficient for turbulent channel flows, which has a similar effect as the Heisenberg eddy viscosity  of Aubry and co-workers. First, we will apply here the constant eddy viscosity--diffusivity, 
but estimate the magnitude of $\eta$ by the method of Cazemier \etal\cite{cazemier} 
as described above. Since 
$Pr=0.7$ and thus $\nu\approx\kappa$, we also use the same amplitudes of $\eta$ for all fields. 
Equations  (\ref{eq:galerkin-proj}) follow to
\begin{align}
\dot{a}^{(p)}_{p_x,p_y} = \sum_{n=1}^{M_Q} {\wp}(p_x,p_y,n,p) a^{(n)}_{p_x,p_y} + (1+ \eta)
\sum_{n=1}^{M_Q} {\epsilon}(p_x,p_y,n,p) a^{(n)}_{p_x,p_y} + \nonumber \\
{N}(p_x,p_y,p) + \sum_{n=1}^{M_Q} \Im (p_x,p_y,n,p) a^{(n)}_{p_x,p_y} + {\epsilon_{\langle T \rangle}}(p_x,p_y,p)
\label{eq:truncation_ODE}
\end{align}
after truncation and addition of an eddy viscosity and diffusivity term 
\begin{equation}
\eta=\alpha \overline{\eta} \,\,\, \text{with} \,\,\, \alpha \in \mathbb{R}\,.
\label{eq:eta_max}
\end{equation} 
The constant eddy viscosity-diffusivity $\overline{\eta}$ is determined by 
\begin{equation}
\overline{\eta} =  \max \left[\frac{{D} (p_x,p_y,p)}{\epsilon(p_x,p_y,n,p)\,\delta_{np}}\right]\,.
\label{eq:definition_eta}
\end{equation}

Figure~\ref{fig:cazemier_aubry} shows the damping term $D_i$ for the POD modes of M2 and 
its ratio with the corresponding dissipation $\epsilon_i$. In these plots $i=(p_x,p_y,p)$ represents the
POD mode $i$ of the LDM. The modes are  ordered by decreasing energy content 
(see also the second column of table \ref{POD_modes}). Note also that in Eq.
(\ref{eq:cazemier_2}) the dissipation term appears only when $n=p$. In the top panel of the figure, the modes are grouped in accordance with their degeneracy as given in Tab.~\ref{POD_modes}, the first two modes are $(0,1,1)$ and $(1,0,1)$ (the other two are obtained from complex conjugation), the second pair $(0,1,2)$ and $(1,0,2)$, etc. 

The order of magnitude of the eddy viscosity and diffusivity term is determined by the maximum 
value of the ratio $D_i/\epsilon_i$, $\overline{\eta} \approx 2.698$ for $i=8$. It is shown in the next section that all the regimes of interest, 
and their typical solutions, can be obtained by a variation of the real prefactor $\alpha$ in the range 
$\alpha \approx 0.87-1.0$ for M1 and $\alpha \approx 0.87-1.04$ for M2.
However, as pointed out by Kalb and Deane \cite{kalb}, this damping term $D_i$ can change sign in contrast to $\epsilon_i$ and can thus add as an additional production term. 

It is also evident from Fig.~\ref{fig:cazemier_aubry} that the Heisenberg eddy viscosity-diffusivity introduces an overwhelming damping for the less energetic modes. This will cause stationary convection 
solutions for both LDMs. These limitations of the model with constant eddy viscosity--diffusivity make it necessary to extend the LDM to a mode--dependent or modal eddy viscosity--diffusivity coefficient, $\eta (i)$. Such a model matches the decreasing ratio $D_i/\epsilon_i$. 
%-----------------------------------------------------------------
\begin{table}
\begin{center}
\begin{tabular}{lcccccc}
\hline\hline
  $k$ & $i-$order  & Modes in LDM  & $\lambda^{(n)}_{n_x,n_y}$  &  $\lambda^{(n)}_{n_x,n_y}/\lambda^{(1)}_{0,1}$ &  Degeneracy & Energy in\%\\
      & in LDM  & $(n_x, n_y, n)$  &    &     &     &    \\
\hline
    1  & 1, 2  & $(0,1,1),(1,0,1) $ &  6.1705  & 1.0000  & 4  &  30.879   \\
    2  & 3, 4  & $(0,1,2),(1,0,2) $ &  1.6710  & 0.2708   & 4  &   8.362   \\
    3  & 5, 6  & $(1,1,1),(1,-1,1)$ &  0.5999  & 0.0972   & 4  &   3.002  \\
    4  & 7, 8  & $(0,2,1),(2,0,1) $ &  0.3788  & 0.0614   & 4  &  1.896 \\
    5  & 9, 10  & $(0,0,1),(0,0,2) $ &  0.3722  & 0.0603   & 2  & 0.932   \\
    6  & 11, 12  & $(0,1,3),(1,0,3) $ &  0.3583  & 0.0581   & 4  &   1.790   \\
    7  & 13, 14  & $(1,2,1),(2,1,1)$ &  0.2729  & 0.0442   & 8  &  2.690    \\
       & 15, 16  & $(1,-2,1),(2,-1,1)$ &       &    &    &        \\
    8  & 17, 18  & $(0,1,4),(1,0,4) $ &  0.2315  &  0.0375  & 4  &  1.158    \\
    9  & 19, 20  & $(0,2,2),(2,0,2) $ &  0.2274  &  0.0369  & 4  &  1.138    \\
   10  & 21, 22  & $(0,3,1),(3,0,1) $ & 0.2146  &  0.0348  & 4  &  1.078     \\
   11  & 23, 24  & $(1,1,2),(1,-1,2) $ & 0.2078  &  0.0337  & 4  &   1.039   \\
   12  & 25, 26  & $(1,1,3),(1,-1,3) $ & 0.1975  &  0.0320  & 4  &   0.971  \\
   13  & 27, 28  & $(2,2,1),(2,-2,1) $ & 0.1630  &  0.0264  & 4  &   0.813   \\
   14  & 29  & $(0,0,3) $ & 0.1554  &  0.0252   &  1  &   0.194   \\
   15  & 30, 31  & $(1,1,4),(1,-1,4) $ & 0.1531  &  0.0248   &  4  &   0.769   \\
\hline\hline
\end{tabular}  
\end{center}
\caption{The first fifteen most energetic POD modes as obtained from the snapshot analysis. The corresponding index in the LDM, the eigenvalues, their ratio with respect to the first eigenvalue, the degeneracy of the modes, and their percentage of the total mean energy content (kinetic energy plus scalar variance) are also given. Ordering is with respect to the eigenvalue $\lambda^{(n)}_{n_x,n_y}$.}
\label{POD_modes}
\end{table}
%---------------------------------------------------------------------------
\begin{figure}[ht!]
\centering
{\includegraphics[width=10.0cm]{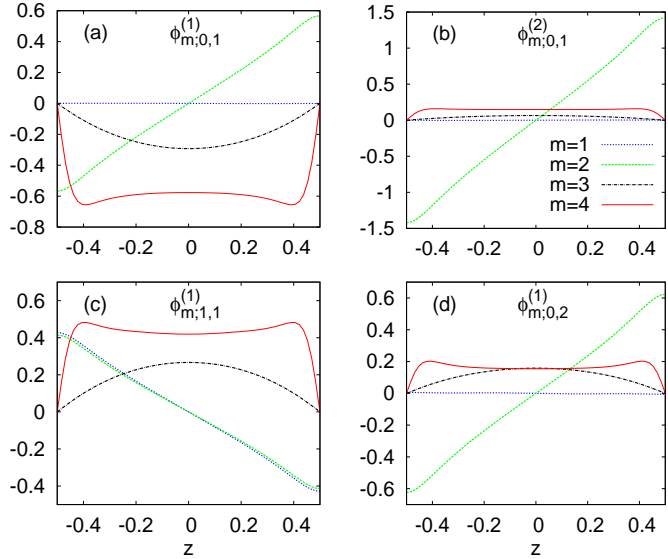}}
\caption{(Color online) The wall-normal dependence of the components of the four most 
energetic  POD modes which is given by $\phi_{m; n_x,n_y}^{(n)}(z)$. The index $m=1,2,3,4$.
The particular mode is indicated in each panel. Linestyles for different components are 
the same in all four subfigures.}
\label{fig:phi_011_111}
\end{figure}
%---------------------------------------------------------------------------

\subsection{Structure of the POD modes}
Before we discuss the time evolution of the LDM with constant and modal eddy 
viscosity--diffusivity, we want to describe the structure of the POD modes. 
Table~\ref{POD_modes} presents the fifteen most energetic modes $\phi^{(n)}_{m;n_x,n_y}$ which are represented by the triplet $(n_x, n_y, n)$. Their decreasing energy content is shown in the fourth and 
fifth column where the individual eigenvalues $\lambda^{(n)}_{n_x,n_y}$ and their ratio with respect 
to $\lambda^{(1)}_{0,1}$ are listed, respectively. In addition, we show their degeneracy and the resulting share in the total energy content. 

The POD analysis has been done for the range of horizontal wavenumbers of $0 \le n_x \le 16$, $-16 \le n_y \le 16$. The order of magnitude of the resulting eigenvalue spectrum is in qualitative 
agreement with  those from Smith \etal~\cite{moehlis_2,moehlis_1} and Moehlis \etal~\cite{moehlis_3} for plane Couette flow, as well as those from Deane and Sirovich~\cite{sirovich_2} for Rayleigh-B\'enard 
convection in a Cartesian cell. The first few modes carry the major share in the total energy and are
followed by a tail of slowly decaying and energetically much less significant degrees of freedom. 
For example, in the plane Couette flow case \cite{moehlis_2,moehlis_1} the first mode carries about 68\%
of the total energy, whereas in the RB convection case \cite{sirovich_2} about 39\% at $Ra \approx 46000$. As shown in Ref. \cite{Bailon2010}, the number of energetically significant modes in RB convection increases rapidly with increasing Rayleigh number. Thus, at $Ra=1.03 \times 10^5$, our primary POD mode only represents about 31\% of the total mean energy. The slow decay of the spectrum results in the  POD mode with the triplet $(0,5,1)$ which is No. $i=101$ in M2 (index $i$ corresponds with the second column of Tab. \ref{POD_modes}) still carrying 0.28\% of the total energy. 

\begin{figure}[ht!]
\centering
{\includegraphics[width=10.0cm]{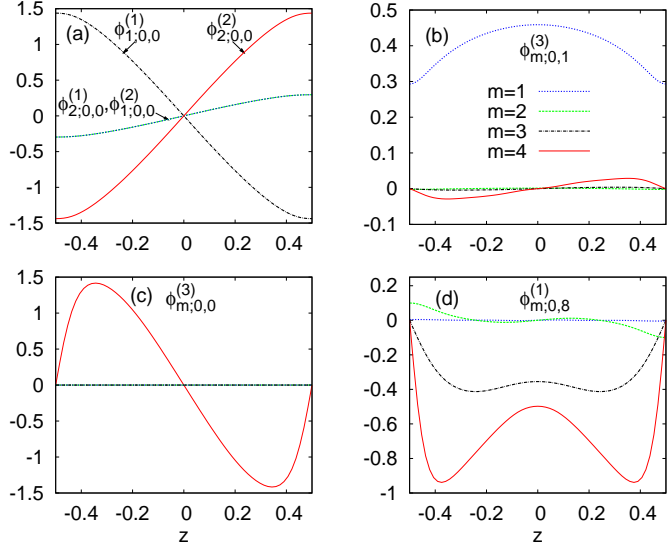}}
\caption{(Color online) Wall-normal dependence of further POD modes. The legend in 
panels (c) and (d) is identical to panel (b).}
\label{fig:phi_001_013}
\end{figure}

Figure~\ref{fig:phi_011_111} displays the wall-normal dependence of the components of the eigenvectors of the four most energetic POD modes, which are  $\phi_{m;0,1}^{(1)}$, 
$\phi_{m;0,1}^{(2)}$, $\phi_{m;1,1}^{(1)}$, and $\phi_{m;0,2}^{(1)}$, all of them representing a 
vertical circulation whose axes, due to their four-fold degeneracy, are along $x$ or $y$, or in 
the case of $\phi_{m;1,1}^{(1)}$ along the horizontal diagonals of the periodic box, respectively. 
We found in our analysis that the convergence of the snapshot method is very slow. This causes
for example the slight differences between $\phi^{(1)}_{1;1,1}$ and $\phi^{(1)}_{2;1,1}$ as seen
in Fig. \ref{fig:phi_011_111}.
Figure~\ref{fig:phi_001_013} presents the wall-normal functionality of further POD modes.
For example, the POD mode belonging to  $\phi_{m;0,0}^{(1)}$ is a purely mechanical 
mode with degeneracy 2 and vanishing components $m=3,4$. The two remaining horizontal components ($m=1,2$), which are denoted as $\phi_{m;0,0}^{(1)}$ and $\phi_{m;0,0}^{(2)}$, satisfy orthogonality (see panel (a)).

Another interesting mode is the sixth most 
energetic POD mode belonging to $\phi_{m;0,1}^{(3)}$.  It represents a pumping motion in 
the $x$ or $y$ directions and is the first of the most energetic POD modes with a non-vanishing 
vertical vorticity component.\cite{tarman_2} The mode is shown in panel (b) of Fig. \ref{fig:phi_001_013}. Furthermore, we also display a purely thermal mode $\phi_{m;0,0}^{(3)}$
with degeneracy 1 in panel (c) and  the mode $\phi_{m;0,8}^{(1)}$ which shows the growing 
influence of the viscous and thermal boundary layers close to the walls at higher horizontal wave numbers in panel (d) of the figure.
%---------------------------------------------------------------------------
%---------------------------------------------------------------------------

\begin{figure}[htpb!]
\centering
\includegraphics[width=7.2cm]{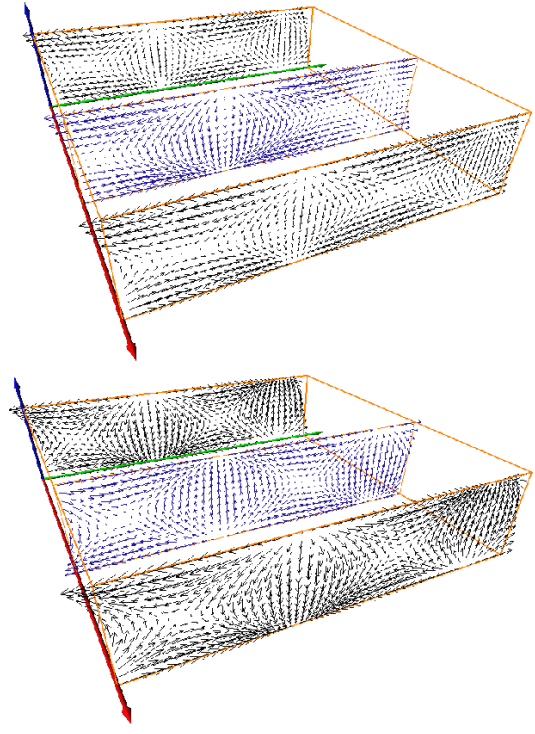}
\caption{(Color online) Vector plots of the first and third most energetic POD modes 
$\Phi_{m;0,1}^{(1)}$ (top) and $\Phi_{m;1,1}^{(1)}$ (bottom).  The velocity field is therefore 
projected into three $y$-$z$ planes.}
\label{fig:Phi_001_111_u}
\end{figure}

\begin{figure}[htpb!]
\centering
\includegraphics[width=7.2cm]{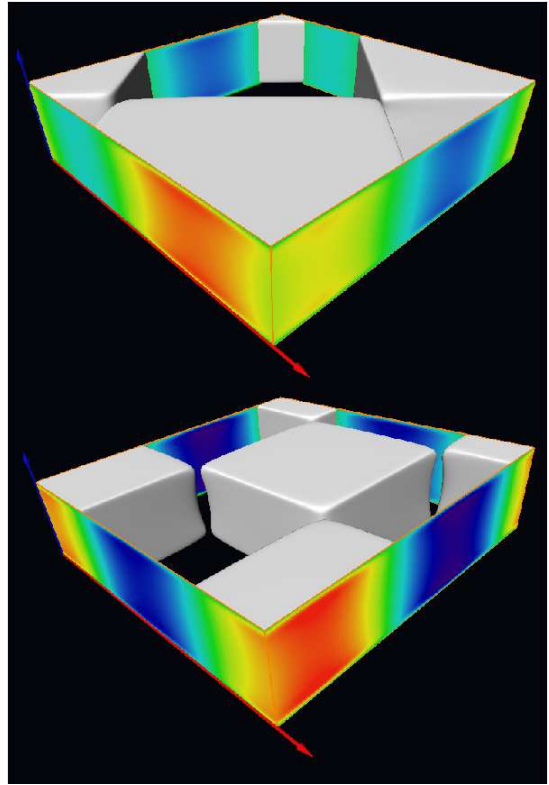}
\caption{(Color online) Isosurfaces of the first and third most energetic POD modes 
$\Phi_{4;0,1)}^{(1)}$ (top) and $\Phi_{4;1,1)}^{(1)}$ (bottom) which describe the temperature fluctuations. The isosurfaces are taken at the level 0.01 $\theta_c$. In addition, we plot 
contours at the sideplanes. Red color stands for 1.7 $\theta_c$ and blue for -1.7 $\theta_c$.}
\label{fig:Phi_001_111_t}
\end{figure}

Figure~\ref{fig:Phi_001_111_u} displays the spatial structure of the velocity field corresponding to 
the functions  $\Phi_{m;0,1}^{(1)}$ and $\Phi_{m;1,1}^{(1)}$. They are obtained  by summing 
all individual eigenfunctions as given by Eq. (\ref{eq:eigenfunc}) over their degeneracy, 
i.e., modes that have the same energy content.
%----------------------------------------------------------------------------
%----------------------------------------------------------------------------
Figure~\ref{fig:Phi_001_111_t} illustrates the  isosurfaces of the temperature fluctuations 
$\theta({\bf x},t)$ that are captured by the fourth component of the same POD modes. 
We show $\Phi_{4;0,1}^{(1)}$ in the top panel and $\Phi_{4;1,1}^{(1)}$ in the bottom panel 
of the figure.
%----------------------------------------------------------------------------
%----------------------------------------------------------------------------

\begin{figure}[htpb!]
\centering
\includegraphics[width=7.2cm]{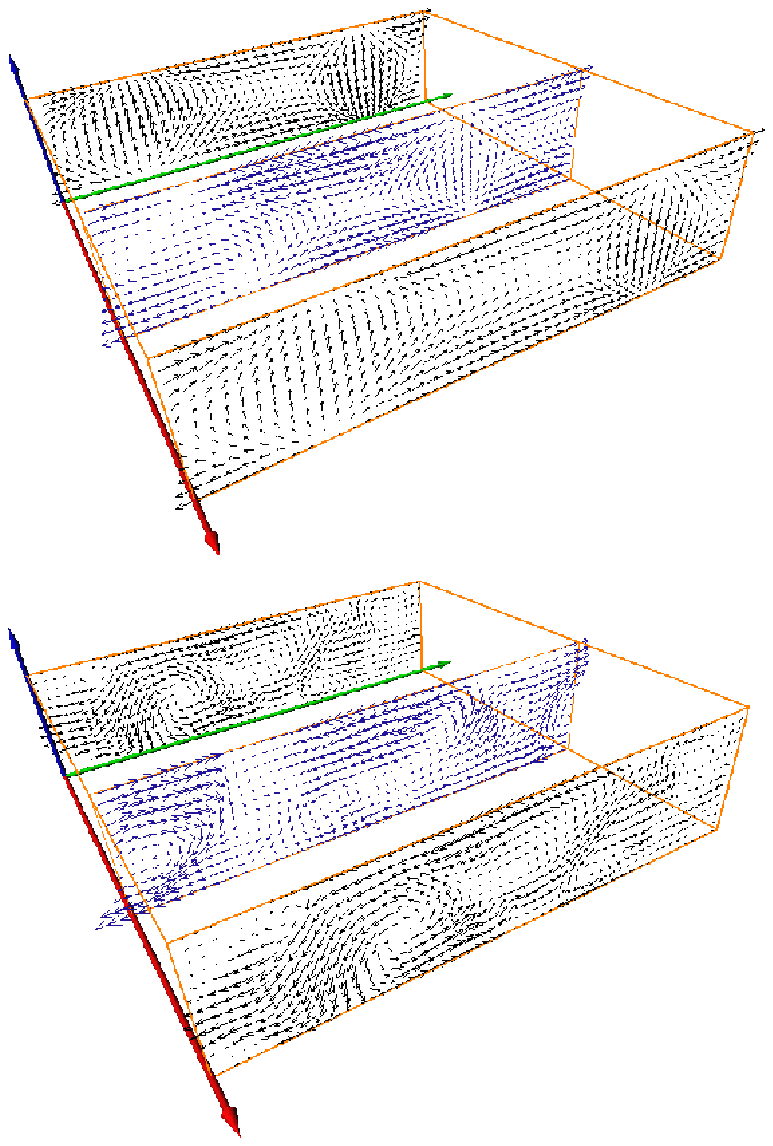}
\caption{(Color online) A snapshot of the velocity field ${\bf u}({\bf x},t_0)$ obtained from the  
DNS (bottom) is reconstructed with the POD modes of low-dimensional model M2 (top).}
\label{fig:u_field}
\end{figure}

\begin{figure}[htpb!]
\centering
\includegraphics[width=7.2cm]{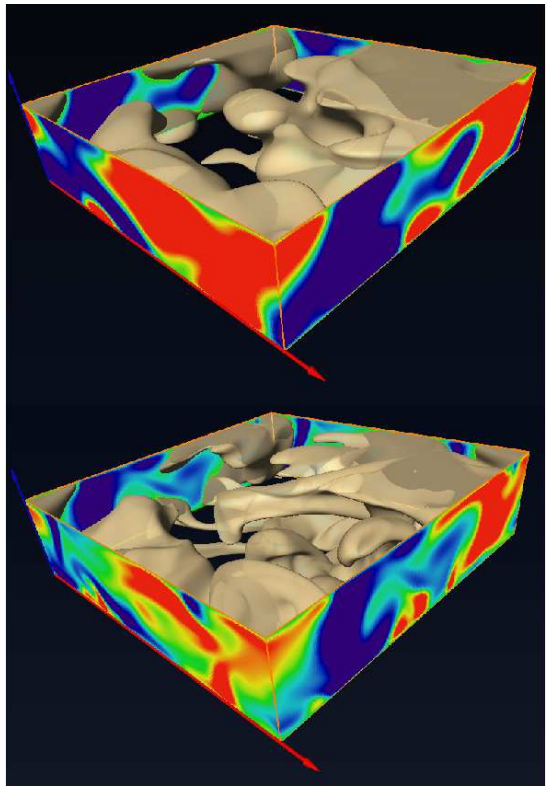}
\caption{(Color online) Temperature fluctuation field $\theta({\bf x},t)$ of a DNS snapshot 
(bottom panel) is reconstructed with the POD modes of low-dimensional model M2 (top panel). 
The isosurfaces are taken at the 
level 0.4 $\theta_c$.  In addition, we plot contours at the sideplanes. Red color stands for 
1.7 $\theta_c$ and blue for -1.7 $\theta_c$.}
\label{fig:t_field}
\end{figure}

One important aspect of the POD mode analysis and LDM setup is the question of how well 
the largest structures and their dynamics can be represented. Here, we display the 
reconstruction of a DNS snapshot by the mode set.  In Fig.~\ref{fig:u_field}, we compare the 
velocity field as reconstructed with the POD modes $\Phi_{m;n_x,n_y}^{(k)}$ from model M2. 
The top panel in the figure shows the reconstruction and the bottom panel the full DNS snapshot.
%----------------------------------------------------------------------------
%----------------------------------------------------------------------------
The same analysis is repeated for the temperature fluctuations at the same instant.
Figure~\ref{fig:t_field} shows the temperature field again reconstructed with M2. Although,
not all details are reproduced, both figures indicate that the most important structures
of temperature and velocity are captured by M2. Our analysis showed that the temperature
fluctuations converged slower than the velocity field. It is known that the temperature field 
forms so called thermal plumes -- fragments of the thermal boundary layer that detach from 
the cooling and heating plates and move into the bulk (see e.g. Refs. 35, 36 and 37). 
These fine-scale filamented structures  carry the heat across the cell.  
%----------------------------------------------------------------------------
%----------------------------------------------------------------------------

\subsection{Time evolution of LDM with constant eddy viscosity--diffusivity}
In the following, we will discuss the evolution of the LDM. This section is for the case with 
the constant eddy viscosity--diffusivity, i.e. $\eta=\alpha \overline{\eta}$ where $\overline{\eta}$ 
is given by~(\ref{eq:definition_eta}), and $\alpha \in \mathbb{R}$. 

\begin{figure}[htpb!]
\centering
\includegraphics[width=8.0cm]{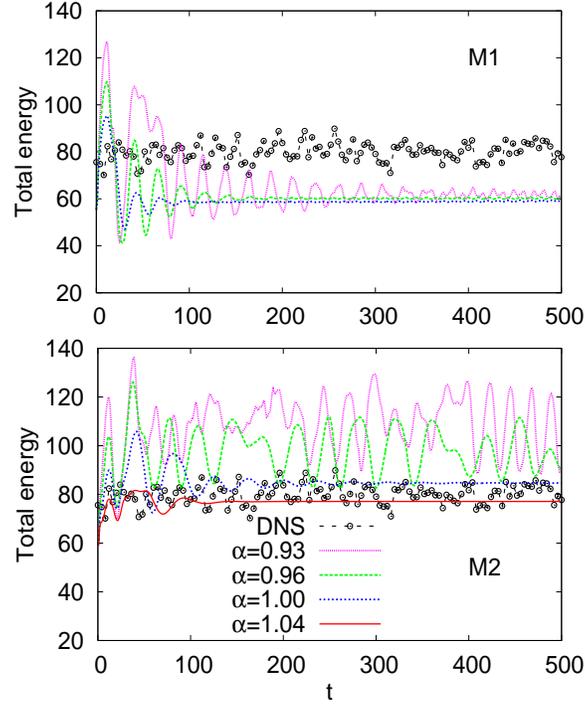}
\caption{Time series of the total energy for LDM M1 (top panel) and M2 (bottom panel). Different magnitudes of the parameter $\alpha$ in the constant Heisenberg eddy viscosity--diffusivity are 
applied in both models and given in the 
legend of the bottom panel. Note that $\alpha=1.04$ is not taken for M1.}
\label{fig:equilibrium}
\end{figure}

As was found in Ref. \cite{cazemier} for the long--term integration of their LDM, this closure 
can lead to a statistical equilibrium state which accumulates too much energy. One
can overcome this behavior in parts following a work by Kalb and Deane \cite{kalb}, who kept the 
original gradient of the mean temperature field, $\frac{d \langle T \rangle}{dz}$, in the 
evaluation of the last two terms of Eq. (\ref{eq:galerkin-proj}). Consequently, the equations 
for $\Im$ and $\epsilon_{\langle T \rangle}$ become
\begin{equation}
\Im(p_x,p_y,n,p)= - \hat{u}_c \int_{-1/2}^{1/2} \frac{d \langle T \rangle}{dz} \phi^{(n)}_{3;p_x,p_y} \phi^{(p)*}_{4;p_x,p_y}\ dz \,.
\label{eq:interaction}\end{equation} 
and,
\begin{equation}
\epsilon_{\langle T \rangle}(p_x,p_y,p) = \sqrt{L_x L_y} \int_{-1/2}^{1/2} \frac{d^2 \langle T \rangle}{dz^2} \phi^{(p)}_{4;0,0} \ dz\,.
\label{eq:mean_diss}
\end{equation}
The long-time evolution of the LDM requires the time integration of the ODE system 
(\ref{eq:truncation_ODE}). A fourth-order Runge-Kutta scheme is applied. Our studies 
found that for $\alpha \lessapprox 0.87$ the ODE systems for M1 and M2 become 
unstable which is triggered by the most energetic modes that accumulate energy which
cannot be transferred sufficiently fast to small scales. For $\alpha \ge 0.87$, we still obtain 
a regime of the LDM carrying too much energy. With increasing prefactor $\alpha$, this 
energy surplus at the first POD modes decreases up to a threshold which enforces the 
whole dynamical system into a stationary state. For  M1 this sets in at $\alpha \ge 1.0$ and 
for M2 at $\alpha \ge 1.04$.  In Fig. \ref{fig:equilibrium}, we compare therefore
the total energy of both LDMs for different values of $\alpha$ with the original DNS data. After a 
relaxation phase both models reach a statistically stationary regime. While model M1
always yields energy time series below that of the original DNS, model M2 comes closer to 
the evolution of the total energy from the original DNS. Note that the fluctuations
of the total energy in both cases are significantly larger than for the original DNS data. This is 
due to the fact that only the largest-scale modes are kept in the model and coupled with each 
other. The exchange of energy among them can cause larger variations. The figure clearly 
indicates that M1 falls short in representing the long-term dynamics of the convective flow. 
In model M2 however it is possible to obtain a  total energy in the range of the original data.   
It can be concluded that a further reduction of degrees of freedom below that of M1 is thus not 
possible.
%--------------------------------------
%--------------------------------------

\begin{figure}[htpb!]
\centering
\includegraphics[width=7.4cm]{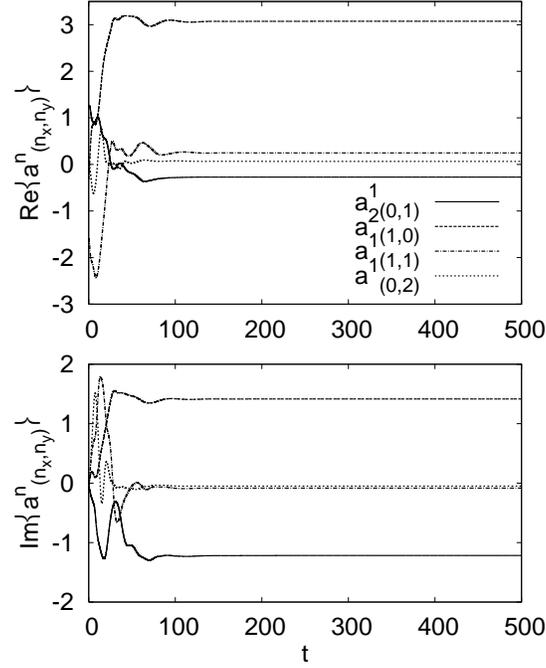}
\caption{Time series of the real and imaginary parts of the expansion coefficients $a_{n_x,n_y}^{(n)}(t)$ of the four most energetic POD modes for M2 and $\alpha =1.04$.}
\label{fig:convergence_DNS_LDM}
\end{figure}

Figure \ref{fig:convergence_DNS_LDM} shows the long-time behavior of $a^{(1)}_{(0,1)}$, 
$a^{(2)}_{(1,0)}$, $a^{(1)}_{(1,1)}$, and $a^{(1)}_{(0,2)}$ for M2 in the stationary overdamped 
state. This fixed point of the LDM is connected with a stationary pattern of the velocity and 
temperature fields. It is now clear how a constant eddy viscosity--diffusivity model produces the observed behavior. The second mode (dashed line) is accumulating too much energy 
($|a^{(2)}_{(1,0)}| > |a^{(1)}_{(0,1)}|$), since it is not damped strongly enough. Similar behavior 
of the LDM has been found and analyzed in detail by Aubry \etal \cite{aubry}. They detected a 
similar fixed point for a specific range of their constant eddy viscosity. 
%--------------------------------------
%--------------------------------------

\begin{figure}[htpb!]
\centering
\includegraphics[width=7.4cm]{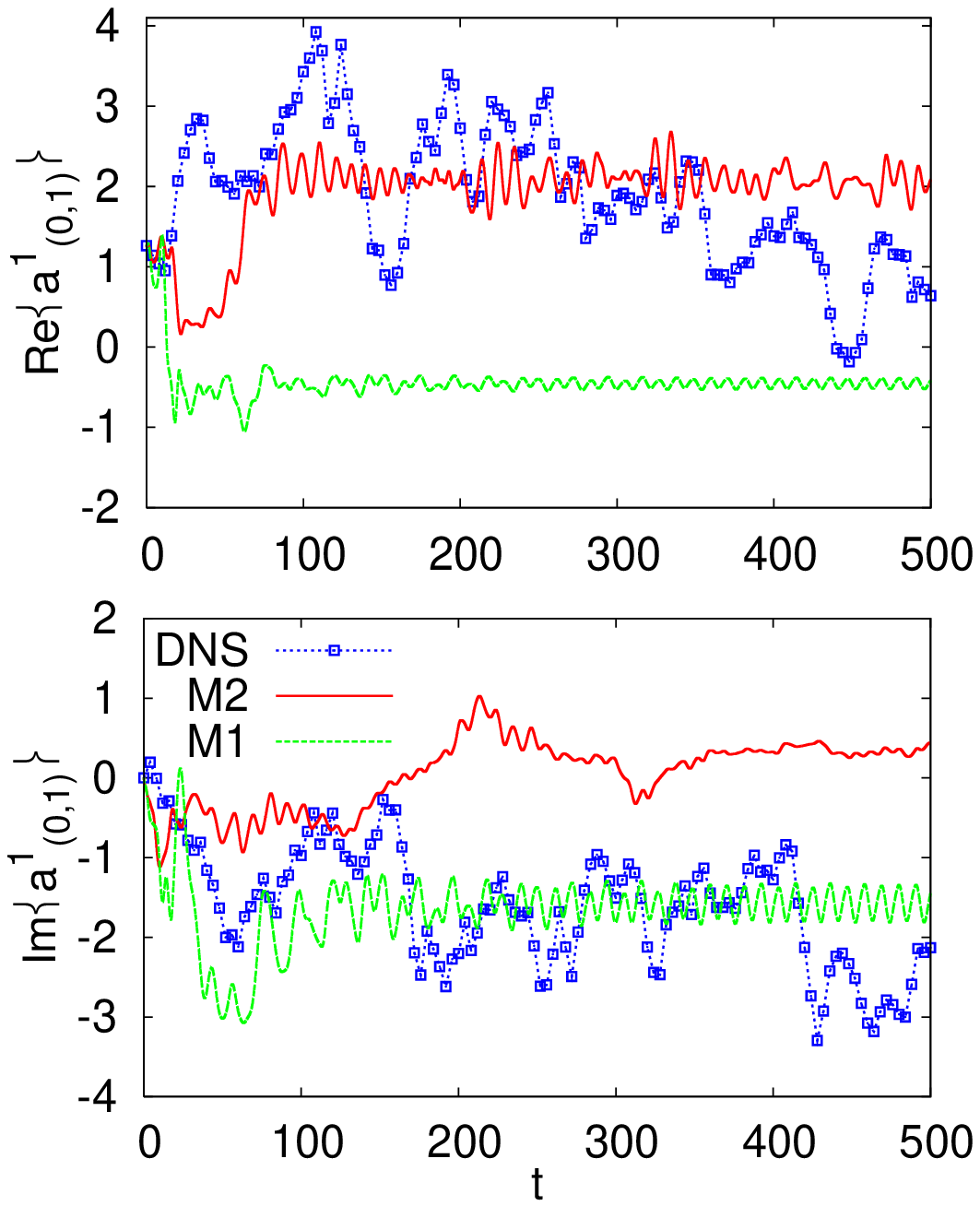}
\caption{Time series for the real (top) and imaginary (bottom) parts of $a^{(1)}_{0,1}$ of the two 
LDMs. The parameter $\alpha$ is set to a value of 0.93.}
\label{fig:at_DNS_LDM}
\end{figure}

Since the overdamped stationary state is not of interest for our study, the values of $\alpha$ 
have to be chosen smaller than these limits. Figure~\ref{fig:at_DNS_LDM} shows the time 
evolution of the modal amplitude for the most energetic POD mode, $a^{(1)}_{0,1}(t)$, obtained 
from the LDMs  M1 and M2, respectively at a smaller value of $\alpha$. Data are again compared 
with DNS time series which are obtained  by projection of the snapshots on the particular modes.  
Clearly, at $\alpha=0.93$ the agreement of the real part is very good for M2, even for long term 
evolution. The amplitude of the imaginary part in M2 is too large, the one in M1 is  comparable with
the DNS.
%--------------------------------------
%--------------------------------------
While the amplitudes partly agree, the temporal behavior of the modes differs qualitatively. 
As observed from Fig.~\ref{fig:at_DNS_LDM},  the real and imaginary parts of the expansion 
coefficients vary in a limited range (in parts periodically) once the initial relaxation
to a statistically stationary state is finished. This is in contrast to the projection of
the DNS snapshots on the modes. It indicates that the additional dissipation has a 
strong impact on the dynamics of the large-scale degrees of freedom. The figure unravels the shortcoming of the present straightforward and simplest closure: all modes are affected by the 
same additional dissipation, the ones that have many couplings within the model as well
as those with much less mode interactions. The constant eddy viscosity--diffusivity establishes 
an additional flux from large resolved to small unresolved scales and seems not to allow
a back-scatter which is important and known from other (subgrid-scale) closures.\cite{Meneveau2000}  

\subsection{Time evolution of the LDM with modal eddy viscosity--diffusivity}
\label{modal_eta}
%--------------------------------------
\begin{figure}[htpb!]
\centering
\includegraphics[width=7.4cm]{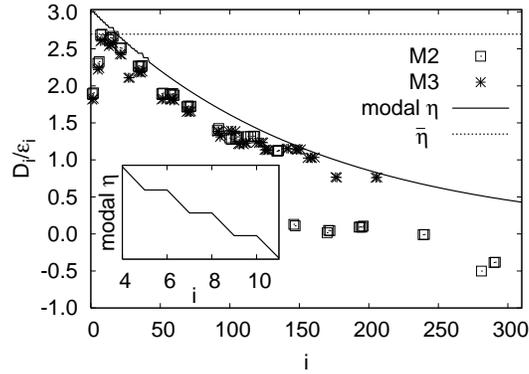}
\caption{Modal eddy viscosity--diffusivity following the maximum values of $D_i/\epsilon_i$. 
The real prefactor is now slightly larger and given by $\alpha =1.146$ for the solid line. 
For comparison, we add the constant value of $\overline{\eta}$ from Fig. \ref{fig:cazemier_aubry}.
Model M3 is used for evaluation of M2 only. The inset shows the degeneracy restrictions that have to be included for the first 41 modes.}
\label{fig:eta_cazemier_max}
\end{figure}
%--------------------------------------
A refinement of the constant eddy viscosity--diffusivity model, that overcomes the shortcomings 
from above, is possible when switching to the modal eddy viscosity--diffusivity. Figure~\ref{fig:eta_cazemier_max} shows the maximum values of the ratio $D_i/\epsilon_i$ that were 
already presented in the top panel of Fig.~\ref{fig:cazemier_aubry} anew for the eddy viscosity--diffusivity of model M2. The sudden decrease in the maxima for $i \approx 140$ is a consequence 
of the truncation. We confirmed this after plotting the modes resulting from a slightly larger
model --denoted as M3-- that contains 430 modes with $|n_x|+|n_y|\le 6$, $1 \le (n) \le 10$ (crossed symbols in the figure). The solid line is a fit which is given by 
\begin{equation}
\eta (i) = \alpha \beta (\gamma)^i \,.
\label{eq:power_law}
\end{equation}
We kept the variable real prefactor $\alpha$ in order to compare this case with the former constant eddy
viscosity--diffusivity. Clearly the new function $\beta (\gamma)^i$ is accounting now for the constant 
$\overline{\eta}$ in the Heisenberg dissipation model. Data fit then well for $\beta = 2.639$, and 
$\gamma = 0.99372$, in the range for $i < 140$. It turned out to be necessary to add some more 
damping through the factor $\alpha > 1$ and to check the results for the ensemble average and the transient total energy.  In agreement with Ref. 8, it yields a positive definite dissipative damping term 
$D_i$ for all modes. The inset  shows the degeneracy for the first 41 modes.
%--------------------------------------
\begin{figure}[htpb!]
\centering
\includegraphics[width=7.4cm]{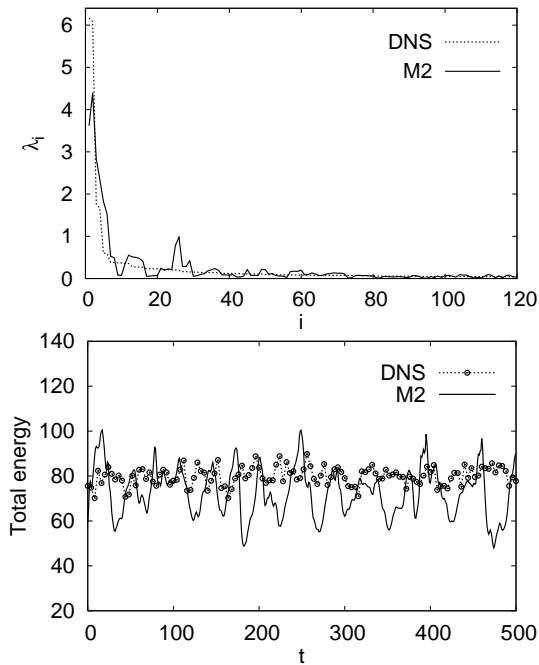}
\caption{Top panel: Mode spectra obtained from  a run of M2 with $\alpha =1.146$ and the original POD of the DNS data. $\lambda_i= \langle a^{p}_{p_x,p_y} a^{p\,\ast}_{p_x,p_y}\rangle$ with $i=(p_x, p_y, p)$ as in section~\ref{truncation_LDM}. Bottom panel:  Time series of the total energies.}
\label{fig:spectrum_cazemier}
\end{figure}
%--------------------------------------

Figure~\ref{fig:spectrum_cazemier} (top panel) displays the energy spectrum of  model M2 after integration of the ODE system following the modal dependent closure model (solid line). It is 
compared with the original POD spectrum of the DNS--data (dashed line). Due to the high accuracy at the tail, the first 120 modes are shown only. It is obvious that a further improvement in the accuracy of the M2--spectrum can be attained considering the effect of the ${\cal G}$--symmetry group (see Appendix~\ref{app_1}) on the time evolution coefficients, as it was done for the full DNS data. The analytical determination of the effect of these symmetries, as was studied by Smith \etal \cite{moehlis_2, moehlis_1} for turbulent plane Couette flow, is beyond the scope of the present study and must be addressed in the future. Differences appear for the two most energetic modes (0,1,1), (1,0,1) which have slightly smaller energy. 
Overshoots for modes No. 12 with $(n_x,n_y,n)=(0,1,3)$ , 25 with (1,-1,3), 26 with (1,1,3), and 29 with (0,0,3) are found. Undershoots are detected for modes No. 9 with (0,0,1), 10 with (0,0,2), 17 with (1,0,4) through 20 with (2,0,2), and 30 with (1,1,4). 

The bottom panel, which is analogous to the bottom panel of Fig.~\ref{fig:equilibrium}, shows the instantaneous total energy for 500 time units. The agreement between DNS and M2 is now
significantly better. Energy remains largely fluctuating and yields an ensemble average equivalent to 91.11\% of the amplitude of the DNS, 5.58\% smaller than the 96.19\% corresponding to constant eddy viscosity--diffusivity at $\alpha=1.04$ (bottom panel of Fig.~\ref{fig:equilibrium}). 
%---------------------------------------------------------------------------
\begin{figure}[ht!]
\centering
{\includegraphics[width=10.0cm]{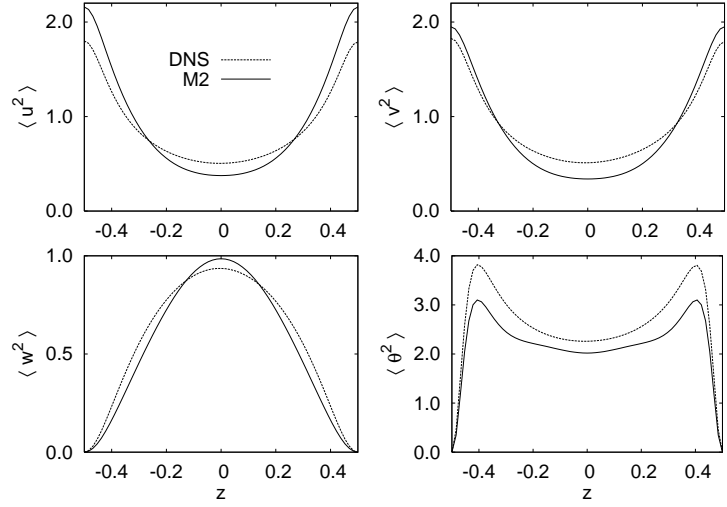}}
\caption{Turbulence statistics $\langle u^2 \rangle$, $\langle v^2 \rangle$, 
$\langle w^2 \rangle$, and ${\langle\theta}^2 \rangle$ versus $z$ as computed from the time
evolution of model M2 and compared with the DNS. Linestyles are the same in the four 
subfigures. The profiles are obtained by taking plane-time averages.}
\label{fig:statistics}
\end{figure}
%---------------------------------------------------------------------------

In Fig.~\ref{fig:statistics} we show the turbulence statistics as obtained from a long-time run of 
model M2 with modal eddy viscosity--diffusivity. Plane and time averaged vertical profiles of 
$\langle u^2 \rangle$, $\langle v^2 \rangle$, $\langle w^2 \rangle$, and ${\langle\theta}^2 \rangle$, computed following equations analog to~(\ref{eq:shear_II}). Time average is taken over the 
first 500 time units. In each case, all the main features of the original DNS--profiles are reproduced, 
and the truncation yields a reasonable accuracy. The profiles are reproduced qualitatively well.   

\begin{figure}[htpb!]
\centering
\includegraphics[width=7.4cm]{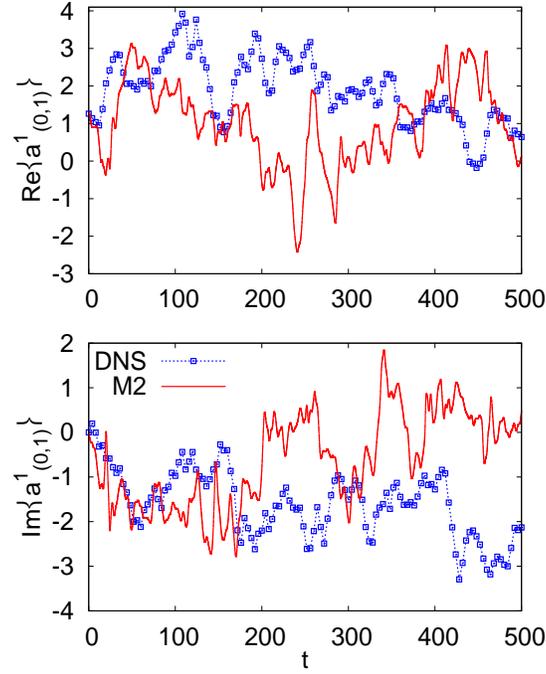}
\caption{Time series for the real and imaginary parts of $a^{(1)}_{0,1}$ of the M2--LDM using the modal--$\eta$ model.}
\label{fig:at_LDM_2}
\end{figure}

Figure~\ref{fig:at_LDM_2} shows the time evolution of the modal amplitude for the most energetic POD mode, $a^{(1)}_{0,1}(t)$, obtained from M2 using the modal eddy viscosity--diffusivity. 
Comparison with the DNS time series shows a reasonable agreement of the real and the imaginary 
parts until about $t=90$. Furthermore, the whole temporal evolution of the coefficients is now much closer to those of the DNS. 
%--------------------------------------
%--------------------------------------
In contrast with Fig.~\ref{fig:at_DNS_LDM}, now the real and imaginary components vary in ranges as wide as the extent of variation for the projection of the DNS snapshots on the modes. The relaxation
into a quasi-periodic time variation as in the case with constant eddy viscosity--diffusivity as now absent. 

\begin{figure}[ht!]
\centering
{\includegraphics[width=12.0cm]{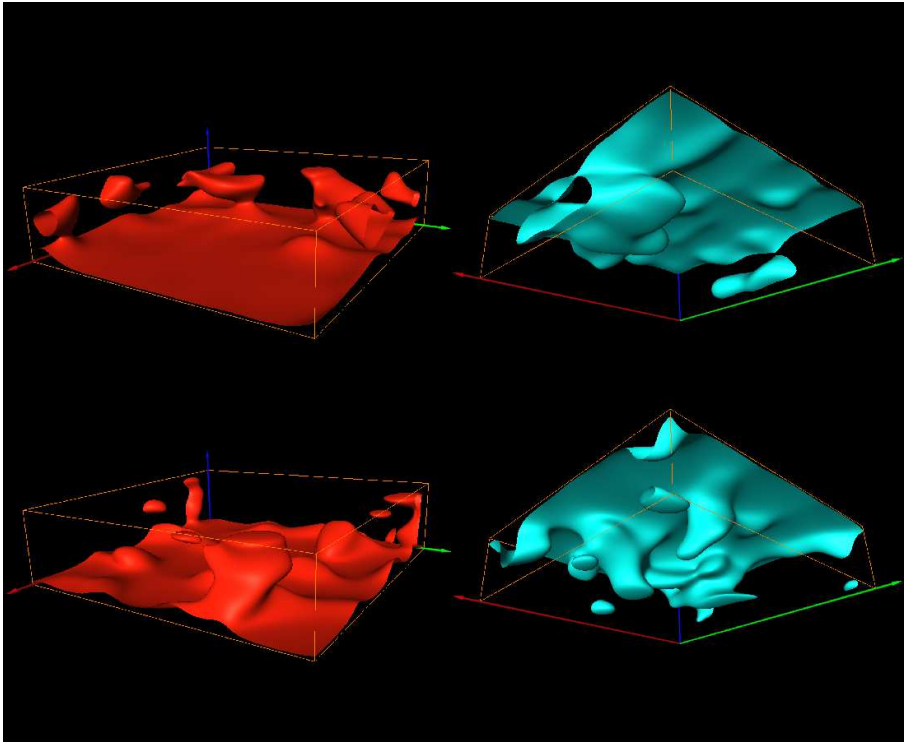}}
\caption{(Color online) Isosurfaces of the total temperature field $\Theta$ at the two instants of time: $t=100$ (top row) 
and $t=400$ (bottom row). Isosurfaces at $\Theta(\mathbf{x},t)/\theta_c = 0.7$ (left column with top view) and $-0.7$ 
(right column with a view from below). Data are obtained from model M2.}
\label{fig:T_plus_theta}
\end{figure}

\begin{figure}[ht!]
\centering
{\includegraphics[width=12.0cm]{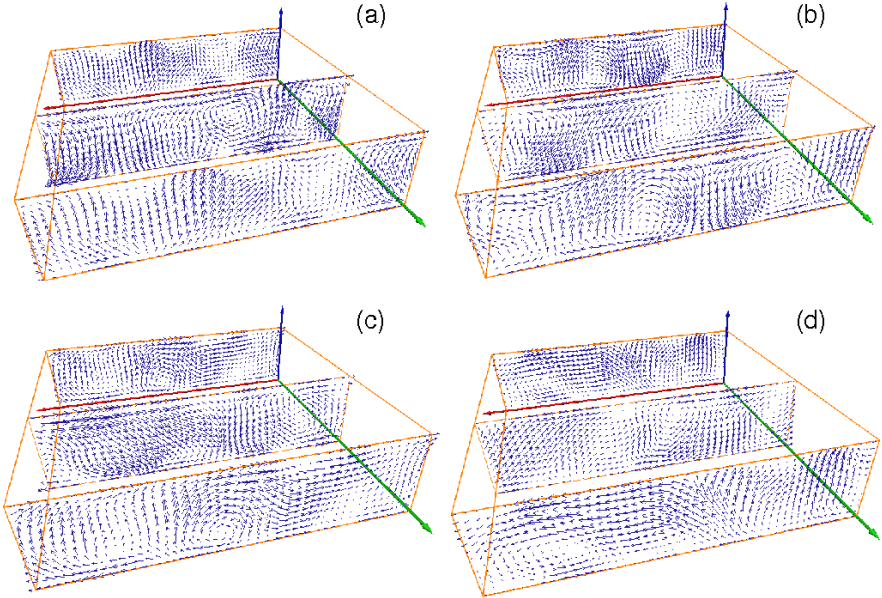}}
\caption{(Color online) Velocity field snapshots at four instants of time: (a) $t=100$, (b) $t=200$, (c) $t=300$, 
(d) $t=400$. Data are obtained from model M2.}
\label{fig:u_LDM_4}
\end{figure}

Figure~\ref{fig:T_plus_theta} shows the total temperature field, $\Theta(\mathbf{x},t)$ 
(see Eq. (\ref{decomposition})), reconstructed with M2, at the two instants of time $t= 100$ and $400$. We clearly identify the mushroom--shaped isosurfaces, which belong to the same type as the ones presented in Fig.~\ref{fig:t_field} for the fluctuations, and were previously observed (see e.g. Refs.~\cite{Shishkina_1,Shishkina_2}) for RB convection in cylindrical containers for $Ra= 10^5$--$10^9$. Velocity snapshots
at the four instants $t= 100, 200, 300$ and $400$ are shown in Fig. \ref{fig:u_LDM_4}. It is observed that the flow is characterized
by strong up- and downward flows. which are in line with the enhanced fluctuations of the total energy in
Fig. \ref{fig:spectrum_cazemier}. 
%---------------------------------------------------------------------------
%---------------------------------------------------------------------------

%---------------------------------------------------------------------------
%---------------------------------------------------------------------------

\section{Summary and outlook}
We have studied a low-dimensional model of turbulent Rayleigh-B\'{e}nard convection in a Cartesian
slab. The POD modes which form the basis of our model have been obtained by a 
so-called snapshot method from a record of 320 statistically independent realizations of a DNS of
convective turbulence for the same geometry. 
Temperature and velocity field fluctuations have to be considered therefore
as a common four-component vector field where the mode selection is done with respect to the 
total energy in the convective flow, i.e. kinetic energy plus thermal variance. The Navier-Stokes-Boussinesq equations are then projected onto the POD modes. The Galerkin projection is truncated 
at two different levels and results in the low-dimensional models denoted as M1 and M2. 

Our results can be summarized as follows. The LDMs have to be stabilized by an additional 
eddy viscosity--diffusivity $\eta$ that assures that the generated energy can be dissipated 
since the small-scale degrees of freedom are missing in the model. This observation is in 
line with existing works on flows in channels or cavities. First, we introduced a constant 
$\eta$ whose order of magnitude, $\overline{\eta}$, was determined as the global maximum quotient 
of the damping term, $D_i$, and the diffusivity $\epsilon_i$.\cite{cazemier} The calculation of 
term $D_i$ is based on the requirement to have a statistically stationary dynamics in the LDM.
We have then studied the dynamics of our LDM as a function of the level of truncation and the magnitude of $\eta$. Similar to the works in simple wall-bounded flow or cavities, we observe a convergence into a stationary regime for amplitudes of $\eta$ that are too large. This regime can 
be considered as a fixed point which is however not in the focus of the present study. 
We showed also that the effect of $\eta$ on the dynamics of the largest-scale modes is to force them into a state with small fluctuations after the passage of a longer transient. 

Alternatively, we introduced a modal eddy viscosity--diffusivity, $\eta (i)$, by fitting the algebraic 
power law of the form $\beta (\gamma)^i$ to all the local maxima of the quotient $D_i/\epsilon_i$, 
and considering the restrictions due to the degeneracy of the most energetic modes. For the model
M2, the long time integration of the ODE system yields solutions with remarkable accuracy in the value of the ensemble average of the total energy (5.6\% below the value for constant $\eta$). Also, the energy spectrum of the POD, is very well reproduced, especially in the tail for modes with higher 
indexes. The vertical profiles of plane-time averaged fluctuations agree qualitatively with those from
DNS. Characteristic coherent structures of convection, such as thermal plumes, are reproduced. 
We can thus conclude that the second approach with the modal eddy viscosity--diffusivity can model 
the long-term dynamics of turbulent convection qualitatively well. We wish to stress here again that this was
in the focus of the present work, namely how far we can advance with a least set of POD modes. 

The question is interesting and important in view to more complex situations, e.g. the problem of 
mixed convection in indoor ventilation systems. Can the same POD framework (with a 
mathematical foundation) be carried 
over to more complex convection flows?  A big advantage of the present turbulent convection case 
in the Cartesian box with periodic side walls is that we have 16 symmetries that significantly 
enhance the data base. In view of applications in more complex geometries, this indicates that 
a similar approach might be much more complicated and could enhance the limitations that 
showed up already for the present turbulent flow. Nevertheless, since the questions, for example with 
the long-term behavior of the large-scale circulation in turbulent convection \cite{Ahlers2009} or indoor
ventilation, are important, we believe that it is still interesting to further follow this route of LDM development based on the POD framework. Some of these efforts will be hopefully presented in the near future.    

\acknowledgements
This work is supported by the Heisenberg Program of the Deutsche Forschungsgemeinschaft 
(DFG). We thank for computing resources on the JUGENE supercomputer at the J\"ulich Supercomputing Centre, J\"ulich (Germany) with grant HIL02 and at the High Performance Computing facility of the University of Puerto Rico. Discussions with B. R. Noack and M. Schlegel are acknowledged.

\appendix
\section{Symmetry considerations}
\label{app_1}
The numerical simulation of Rayleigh-B\'enard convection for the case of a square section, $L_x=L_y$, generates a maximal amount of symmetry. These discrete symmetries form a group $\mathbf{G}$ of eight elements \cite{sirovich_5}
\begin{equation}
\mathbf{G}=\{I,R,R^2,R^3,F,FR,FR^2,FR^3\}
\label{eq:symmetry}
\end{equation}
whose generators are the rotation by $90^\circ$
\begin{equation}
\mathbf{R}(x,y,z,u,v,w,\theta)=(-y,x,z,-v,u,w,\theta)
\label{eq:rotation}
\end{equation}
and the reflection in $x$
\begin{equation}
\mathbf{F}(x,y,z,u,v,w,\theta)=(-x,y,z,-u,v,w,\theta)
\label{eq:reflection}
\end{equation}
Furthermore, another symmetry group $\{\mathbf{I},\mathbf{Z}\}$ acts on the vertical direction, where $\mathbf{Z}$ is the reflection in $z$
\begin{equation}
\mathbf{Z}(x,y,z,u,v,w,\theta)=(x,y,-z,u,v,-w,-\theta)
\label{eq:y-reflection}
\end{equation}
resulting in a symmetry group of sixteen elements
\begin{equation}
{\cal G}=\{\mathbf{G}, \mathbf{Z} \mathbf{G}\}
\label{eq:symmetry-total}
\end{equation}
when combined. Since each element of the symmetry group generates a possible flow, the ensemble is enlarged by a factor of sixteen, thus increasing the accuracy of any statistical evaluation of the flow.

\end{document}